\documentclass[nohyper]{JHEP3}
\usepackage{graphicx}
\usepackage{amsmath}
\usepackage{amssymb}

\preprint{
NIKHEF/2009-013\\
CERN-TH/2009-136\\
ITF-UU-09/31\\
CP3-09-33
}

\title{Isolating $Wt$ production at the LHC}

\author{Chris D. White\\
  Nikhef Theory Group,  Science Park 105, 1098 XG Amsterdam, The Netherlands\\
  E-mail: \email{cwhite@nikhef.nl}}

\author{Stefano Frixione%
  \thanks{On leave of absence from INFN, Sez. di Genova, Italy}\\
 PH Department, Theory group, CERN 1211-CH Geneva, Switzerland \\
 ITPP, EPFL, CH-1015 Lausanne, Switzerland\\
 E-mail: \email{Stefano.Frixione@cern.ch}}

\author{Eric Laenen\\
  ITFA, University of Amsterdam,
  Valckenierstraat 65, 1018 XE Amsterdam \\
  ITF, Utrecht University, Leuvenlaan 4, 3584 CE Utrecht\\
  Nikhef Theory Group,  Science Park 105, 1098 XG Amsterdam, The Netherlands\\
  E-mail: \email{Eric.Laenen@nikhef.nl}}

\author{Fabio Maltoni\\
Center for Particle Physics and Phenomenology (CP3), Universit\'{e} catholique de Louvain, B-1348 Louvain-la-Neuve, Belgium\\
E-mail: \email{fabio.maltoni@uclouvain.be}}

\abstract{
We address the issue of single top production in association with a $W$ boson at the Large Hadron Collider, in particular how to obtain an accurate description in the face of the top pair production background given that the two processes interfere with each other. We stress the advantages of an MC@NLO description, and find that for cuts used to isolate the signal, it makes sense to consider $Wt$ as a well-defined production process in that the interference with $t\bar{t}$ production is small, and the cross-section of the former is above the scale variation uncertainty associated with the latter. We also consider the case where both $Wt$ and $t\bar{t}$ production are backgrounds to a third process (Higgs boson production followed by decay to a $W$ boson pair), and find in this context that interference issues can also be neglected. We discuss the generalization of our results to other situations, aided by a comparison between the MC@NLO approach and a calculation of the $WWb\bar{b}$ final state matched to a parton shower.
}

\begin{document}
\section{Introduction}
\label{intro}
Top physics is an active research area, not least because the mass of the top quark is close to the scale of electroweak symmetry breaking. Given that one expects theories beyond the Standard Model (SM) to explain this symmetry breaking, it follows that the top sector is a potentially sensitive probe of new physics effects. Top quark production is also of interest within the SM, for precision measurements of masses and couplings, and as a background to other processes. \\

\begin{figure}[h]
\begin{center}
\scalebox{1.0}{\includegraphics{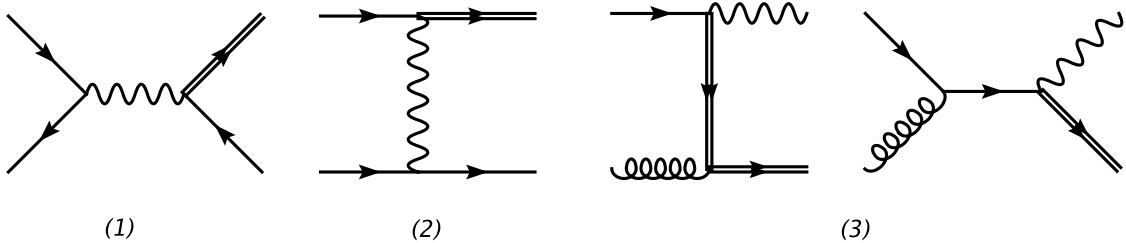}}
\caption{The three SM single top production modes, shown at LO: (1) $s$-channel production; (2) $t$-channel production; (3) $Wt$ production. Double lines represent the top quark.}
\label{modes}
\end{center}
\end{figure}

Single top physics (in which a $t$ or $\bar{t}$ is produced without its accompanying antiparticle) is of particular interest, given that the LO processes are all purely electroweak in nature. The corresponding diagrams are shown in figure \ref{modes}, and there are three distinct production modes. 
The first two are conventionally referred to as the $s$- and $t$-channel modes (depending on the nature of the exchanged $W$ boson), and have been recently identified (in combination) at the Tevatron~\cite{Aaltonen:2009jj,Abazov:2009ii}. The third mode is that of $Wt$ production, and is distinguished by the presence of a $W$ boson accompanying the single top quark in the final state. Its cross-section is rather too small to be observed at the Tevatron, but makes up about 20\% of the total single top cross-section at the LHC, whilst the $s$-channel mode becomes negligible. \\

It is desirable to isolate $Wt$ production for a number of reasons. Firstly, it is sensitive to new physics effects which modify the $Wtb$ vertex of the Standard Model, but not to effective 4-fermion interactions (which mainly affect the $s$- and $t$-channel modes). Thus, it is in principle a different test of BSM theories (see e.g.~\cite{Cao:2007ea} for a model-independent analysis). Secondly, it offers complementary information on the $Wtb$ vertex within the Standard Model (e.g. the value of the CKM matrix element $V_{tb}$ in connection with the possibility of a fourth generation~\cite{Alwall:2006bx,Kribs:2007nz,Soni:2008bc,Chanowitz:2009mz,Holdom:2009rf}). Furthermore, $Wt$ production is a background to many processes, including both neutral and charged Higgs boson production. In such cases one must evaluate the sum of top pair production and $Wt$ production as a background, and it is important that this be done consistently. \\

The cross-sections for single top production in the $s$- and $t$-channel modes have been calculated at NLO in QCD in~\cite{Harris:2002md,Sullivan:2005ar,Sullivan:2004ie,Campbell:2004ch}, with decay effects studied in~\cite{Cao:2004ky,Cao:2004ap,Cao:2005pq}. Recently, the $t$-channel mode was calculated at NLO in the four-flavor scheme, in which initial state $b$ quarks are generated from gluon splitting~\cite{Campbell:2009ss}. The $Wt$ cross-section was first considered in~\cite{Tait:1999cf}, and has also been calculated at NLO in QCD~\cite{Zhu:2002uj,Campbell:2005bb}. Furthermore all three production modes have been implemented in the MC@NLO software framework for combining NLO matrix elements with a parton shower algorithm~\cite{Frixione:2002ik,Frixione:2005vw,Frixione:2008yi}, including spin correlations in the top decay products using the method outlined in~\cite{Frixione:2007zp}\footnote{For a recent study of spin correlations in single top production, see~\cite{Motylinski:2009kt}.}. This constitutes the state of the art for the description of single top physics\footnote{The $s$- and $t$-channel processes at NLO were very recently interfaced with a parton shower in the POWHEG framework~\cite{Alioli:2009je}.}, combining the reduction of theoretical systematic uncertainties which result from adopting an NLO description of the hard event with the high multiplicity, hadron-level events resulting from the parton shower algorithm. The latter can furthermore be interfaced with detector simulations.\\

The calculation of the $Wt$ mode at NLO is non-trivial (and its implementation in MC@NLO is no exception), as discussed in~\cite{Frixione:2008yi}, due to the fact that the $Wt$ production process (at NLO) interferes with $t\bar{t}$ production (at LO), with decay of the $\bar{t}$ (or $t$ quark in the case of $W\bar{t}$ production). It becomes unclear whether it is meaningful to define $Wt$ production as a separate signal in its own right, or whether one should instead consider combining $Wt$ and $t\bar{t}$ production, i.e. only consider given final states comprised of $W$ bosons (or their decay products) and $b$ quarks. The latter approach has practical problems of its own, and the question arises of how to obtain the theoretically most accurate description of $Wt$ production. In~\cite{Frixione:2008yi} two definitions of the $Wt$ mode were given, such that the difference between them measures the interference between $Wt$ and $t\bar{t}$ production. This interference is not guaranteed to be small over all of phase space, but by comparing the results obtained from the two codes it is possible to ascertain whether or not it makes sense to be considering $Wt$ production as an independent process. This problem is not explicitly encountered in previous analyses of the $Wt$ mode by experimental collaborations, which use LO Monte Carlo descriptions (based on the five flavor scheme, in which $b$ quarks are present in the initial state).\\

The aim of this paper is to further investigate these questions, and to investigate various strategies of how to theoretically describe the $Wt$ mode. There are two issues to consider: the reduction of interference between $Wt$ and $t\bar{t}$ production (i.e. to what extent the former is well-defined), and furthermore whether $Wt$ can be efficiently isolated as a signal or reduced as a background. The answer to both of these questions depends on the experimental cuts applied. However, they are related issues in the sense that cuts used to isolate the $Wt$ signal will also influence the interference between $Wt$ and $t\bar{t}$ production.\\

The paper is organized as follows. In the next section, we recall the interference problem between the $Wt$ and $t\bar{t}$ production processes. In section \ref{signal} we consider the isolation of $Wt$ production as a signal, and show that for fairly loose cuts the $Wt$ cross-section is visible above the scale dependence of the $t\bar{t}$ background, and that interference between the two processes is small. In section \ref{Higgs} we consider the case of $Wt$ production as a background to a third process, that of a Higgs boson decaying to a $b\bar{b}$ pair, and show that in this case interference effects are also small, such that one may consider $Wt$ and $t\bar{t}$ production as distinct background processes. In section \ref{finalstates} we examine another approach for describing $Wt$ production, namely that of consistently combining $Wt$ and $t\bar{t}$-like diagrams, and consider the relative merits with respect to the MC@NLO calculation. We discuss our results in section \ref{discussion} and conclude.
\section{Interference problem}
\label{inter}
\begin{figure}[h]
\begin{center}
\scalebox{1.0}{\includegraphics{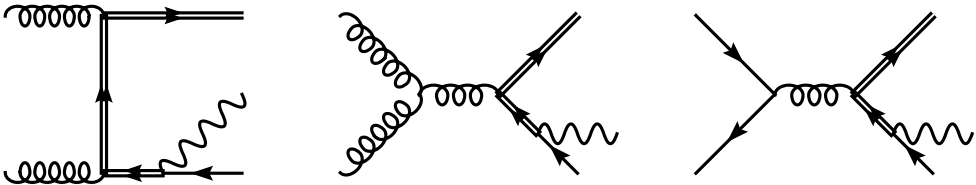}}
\caption{A subset of diagrams contributing to $Wt$ production at NLO, consisting of top pair production, with weak decay of one of the final state top particles.}
\label{NLOdiags}
\end{center}
\end{figure}

At NLO in QCD, the $Wt$ mode (shown at LO in figure \ref{modes}) includes the corrections shown in figure \ref{NLOdiags}. 
Such diagrams can also be thought of as the production of a top quark pair, with decay of the $\bar{t}$ (or $t$ quark in the case of single antitop production in association with a $W$ boson). A problem then occurs if the invariant mass of the final state $Wb$ system is close to the top mass, in that the propagator for the intermediate top particle becomes large. More specifically, the $Wt$ and $t\bar{t}$ cross-sections are well-defined at LO, with $\sigma_{Wt}<\sigma_{t\bar{t}}$. The NLO correction to $Wt$, including the diagrams shown in figure \ref{NLOdiags}, then represents a huge correction, effectively undermining the perturbative description of the $Wt$ mode. There are two main viewpoints for how to deal with this problem.\\

The first, and at first sight the most theoretically rigorous approach, is to conclude that $Wt$ production does not exist, and that its status as an independent production process is an accident of perturbation theory at leading order. One then considers given final states, and sums all possible Feynman diagrams to a given order in $\alpha_S$ and $\alpha_{EW}$ which lead to those final states. In this case the relevant final states are $WWb$ and $WWbb$, i.e. where $b$ may denote (anti-)bottom quarks as appropriate\footnote{This assumes a calculational framework in which initial state $b$ quarks are present (i.e. a five flavor number scheme for the parton densities). The discussion is modified in a four flavor scheme, in which all $b$ quarks are generated explicitly from gluon splittings, as we will see later in the paper.}, and $WW$ denotes $W^+W^-$. Disregarding other backgrounds, the $WWb$ state receives contributions from LO $Wt$ production (as depicted in figure \ref{modes}, following decay of the top), whereas $WWbb$ receives contributions from NLO $Wt$ graphs as well as LO $t\bar{t}$ graphs. However, the use of the terms $Wt$ (or $t\bar{t}$) production does not really make sense in this viewpoint, as only given final states are physically meaningful. Although this approach naturally incorporates interference effects, it suffers from severe phenomenological and technical problems in practice. In particular, corrections to the $WWbb$ final state arising from NLO QCD contributions to $t\bar{t}$ production (followed by decay of both top particles) have not been computed in the above superposition of $Wt$ and $t\bar{t}$. However, these corrections are known to be large for $t\bar{t}$ production (as we will see), significantly limiting the accuracy of the description if they are not included.\\ 

There are also practical reasons why separation of $Wt$ and $t\bar{t}$ production is useful. If one is trying to isolate single top production as a signal, one wishes to efficiently obtain samples of Monte Carlo events corresponding to this signal. If one only has a tool for generating the combination of single and top pair production, most of the generated events will fail the signal cuts, such that event generation efficiency for the $WWb$ final state is low. \\

These problems motivate a second viewpoint, namely that one is allowed to consider $Wt$ as a well-defined process, {\it subject to adequate cuts}. This relies upon the observation that when cuts are applied to isolate the $WWb$ final state, interference effects may be small in practice. Thus one may consider them, for practical purposes, as arising from $t\bar{t}$ production with no subsequent interference between single top and top pair production. To be more specific, let us split the full NLO corrections to the LO $Wt$ amplitude into two parts as follows:
\begin{equation}
{\cal A}_{Wt}= {\cal A}_1+{\cal A}_2,
\label{ampsplit}
\end{equation}
where the first term on the right-hand side contains diagrams with only one top quark (either real or virtual), and the second term corresponds to the diagrams in figure \ref{NLOdiags} containing two top particles in an intermediate state. The squared amplitude is then given by
\begin{equation}
|{\cal A}_{Wt}|^2\propto|{\cal A}_1|^2+2\text{Re}[{\cal A}_1^\dag {\cal A}_2]+|{\cal A}_2|^2.
\label{ampsplit2}
\end{equation}
One can choose to interpret the first term to be a part of the $Wt$ production process (which has a well-defined NLO QCD correction), and the third term to be due to LO $t\bar{t}$ production. This interpretation is only meaningful provided the interference term $2\text{Re}[{\cal A}_1^\dag {\cal A}_2]$ is small, and whether or not this is the case depends strongly on the cuts applied. We will see later in the paper that cuts that are typically used to isolate the $Wt$ mode at LO do reduce the interference term occurring at NLO, and thus the notion of a $Wt$ production process with a well-defined NLO correction does indeed make sense. If such cuts are used, the process of consistently considering the $Wt$ signal plus $t\bar{t}$ background then amounts to generating separately samples of $Wt$ and $t\bar{t}$ events (at LO or NLO as desired) and adding together the results. Similar considerations apply if $Wt$ and $t\bar{t}$ production are both backgrounds to a third production process, provided the isolation cuts associated with the third process are such as to render the interference between $Wt$ and $t\bar{t}$ small. The advantages of such an approach are obvious:
\begin{itemize}
\item One can efficiently generate both $Wt$ and $t\bar{t}$ events up to NLO for use in an analysis, of particular advantage when $Wt$ is the signal.
\item NLO corrections can be included in both processes i.e. one has separate $K$-factors for each, greatly increasing the theoretical accuracy of the description.
\item Previous analyses of $Wt$ production at LO can also be consistently performed at NLO, provided (as is indeed usually the case) that the LO cuts reduce the interference term with $t\bar{t}$ production.
\end{itemize}

The idea of $Wt$ production as a well-defined process at NLO is not new. Indeed, every previous calculation of $Wt$ production beyond LO (including those analyses which only include tree level diagrams) has had to define some prescription for dealing with the interference problem~\cite{Zhu:2002uj,Campbell:2005bb}. These approaches were compared in detail in~\cite{Frixione:2008yi}, and we do not repeat the discussion here. Also in~\cite{Frixione:2008yi}, two definitions of $Wt$ production were given in the context of a full parton shower approach at NLO. These definitions were called {\it diagram removal} (DR) and {\it diagram subtraction} (DS), where the former removes resonant $t\bar{t}$ effects from $Wt$ at the amplitude level (by not including the diagrams of figure \ref{NLOdiags}), and the latter at the cross-section level. The difference then in essence measures the interference between $t\bar{t}$ and $Wt$ production\footnote{The reader may worry about violation of gauge invariance. This is discussed at length in~\cite{Frixione:2008yi}.}. Furthermore, both of these definitions are implemented in the MC@NLO event generator (see~\cite{Frixione:2008ym} for technical information). By running the same analysis with both the DR and DS codes, one is able to check whether interference effects are a problem for a given set of analysis cuts, or not.\\

If indeed the interference has been shown to be small, then one has succeeded in separating the signal plus background of $WWb$ and $WWbb$ final states into two non-overlapping parts, which we may call $Wt$-like and $t\bar{t}$-like signatures. This separation of the final states is, as stated clearly above, dependent on cuts. Where such cuts are used, however, the $Wt$ and $t\bar{t}$ separation is a very good (and, importantly,  quantifiable) approximation to the underlying physics.\\

Successful isolation of the $Wt$ mode requires not only that the interference with $t\bar{t}$ is reduced, but also that a good signal to background ratio can be obtained. For example, it has not yet been shown whether the size of the $Wt$ cross-section is such that it can be significantly observed relative to the systematic uncertainty associated with the $t\bar{t}$ background. This is the subject of the following section.
\section{Isolating the $Wt$ signal}
\label{signal}
In this section we investigate whether it is meaningful to describe a signal of $Wt$ production above a background of $t\bar{t}$ production. We require two criteria to be satisfied. Firstly, that the interference between $Wt$ and $t\bar{t}$ production can be neglected, as can be checked by comparing results obtained with DR and DS. Secondly, that the $Wt$ cross-section is larger than the scale variation associated with the $t\bar{t}$ result. The latter is an indication of whether the identification of $Wt$ is meaningful given the systematic errors associated with the (potentially large) background, and will not be satisfied for generic cuts. \\

Given that we are only considering interference aspects of $Wt$ production in this paper, we neglect all backgrounds apart from top pair production. In more realistic analyses, further cuts should be applied, but one does not expect these to weaken any separation of $Wt$ and $t\bar{t}$ that has been achieved with looser cuts. Motivated by previous studies (e.g.~\cite{Aad:2009wy}), we consider the following cuts:
\begin{center}
\textbf{$Wt$ signal cuts}
\end{center}
\begin{enumerate}
\item The presence of exactly 1 $b$ jet with $p_T>50$ GeV and $|\eta|<2.5$. No other $b$ jets with $p_T>25$ GeV and $|\eta|<2.5$.
\item The presence of exactly 2 light flavor jets with $p_T>25$ GeV and $|\eta|<2.5$. In addition, their invariant mass should satisfy $55$ GeV$<m_{j_1j_2}<85$ GeV.
\item The presence of exactly 1 isolated lepton with $p_T>25$ GeV and $|\eta|<2.5$. The lepton should satisfy $\Delta R>0.4$ with respect to the two light jets and the $b$ jet, where $R$ is the distance in the $(\eta,\phi)$ plane.
\item The missing transverse energy should satisfy $E^{miss}_{T}>$25 GeV. 
\end{enumerate} 
These cuts are designed to isolate semileptonic decays of the two $W$ bosons, one of which comes from the decay of the top quark in $Wt$. These are cleaner than fully hadronic decays (due to backgrounds), but with a cross-section sizeable enough so that studies are possible with early LHC data. Preference for the semi-leptonic decay mode comes from the presence of the isolated lepton, and the missing transverse energy requirement (stemming from the presence of a neutrino in the final state). Moreover, one expects most $Wt$-like events to have only one hard $b$ jet whereas $t\bar{t}$ events have two $b$ jets at LO parton level. Hence, the requirement of exactly one hard $b$ jet in the final state significantly reduces the $t\bar{t}$ background, and also (as we shall see) the interference between $Wt$ and $t\bar{t}$ production. The latter is not surprising, as it has already been shown that a transverse momentum veto on the second hardest $b$ jet reduces very efficiently the interference between single top and top pair production~\cite{Campbell:2005bb,Frixione:2008yi}. A cut on the number of $b$ jets of given $p_T$ is clearly closely related to the notion of a veto on additional $b$ jets. In practice, there will be a number of $Wt$-like events due to $t\bar{t}$ production, where one of the $b$ jets in $t\bar{t}$ is either too soft to be detected, or has been misidentified as a light jet.\\

In order to model such effects, we apply the above cuts for a number of choices of $b$ tagging efficiency $e_b$ and light jet rejection rate $r_{lj}$. That is, $b$ jets are kept with a probability $e_b$, and otherwise taken to be light jets. Similarly, light jets are kept with a probability $1-1/r_{lj}$ (using the conventional definition of the rejection rate), and otherwise taken to be $b$ jets. We assume the same efficiencies for every jet. This may not be the most realistic model, but the hope is that considering different values for $e_b$ and $r_{lj}$ adequately explores the systematic uncertainty due to these effects. The choices can be found in table \ref{DRvsDS}. We also show results with $e_b=1$ and $r_{lj}=10^4$, i.e. a default Monte Carlo calculation without $b$ tagging effects or light jet rejection included.\\

The cut on the invariant mass of the light jet pair helps to discriminate both $Wt$ and $t\bar{t}$ production from other backgrounds. However, it also helps reduce $t\bar{t}$ relative to $Wt$ production, as it requires that the invariant mass of the light jet pair lies within a window of the $W$ mass i.e. that the two light jets result from the decay of a $W$ boson. Given that there are more jets on average in top pair production, the chance that the two jets entering the cuts have both arisen from the same $W$ boson is smaller.\\

The above cuts are reasonably loose, particularly given that most $p_T$ and $\eta$ cuts arise from detector constraints. Extra cuts would in practice be used to tighten the signal to background ratio. However, our aim here is merely to show that even for cuts that are not particularly strict, a clean separation of $Wt$ and $t\bar{t}$ production can be found. Additional cuts aimed at enhancing the signal should then further reduce the interference.\\

The cross-sections that result after application of the above cuts are shown in table~\ref{DRvsDS}. All results have been obtained using a top mass and width of $m_t=170.9$ GeV and $\Gamma_t=1.4$ GeV respectively. The $W$ mass and width are $M_W=80.42$ GeV and $\Gamma_W=2.141$ GeV. We use the MRST2002 NLO parton densities~\cite{Martin:2002dr}. By default, renormalization and factorization scales are set to $\mu_F=\mu_R=m_t$. 
\begin{table}
\begin{center}
\begin{tabular}{c|c|c|c|c}
$e_b$&$r_{lj}$&$\sigma^{\text{DR}}_{Wt}$/pb&$\sigma^{\text{DS}}_{Wt}$/pb&$\sigma_{t\bar{t}}$/pb\\\hline
1.0&$10^4$& $1.206^{+0.039}_{-0.017}$ & $1.189^{+0.021}_{-0.010}$ &$5.61^{+0.74}_{-0.54}$\\
0.6&30& $0.717^{+0.020}_{-0.014}$  &  $0.696^{+0.020}_{-0.005}$ & $4.29^{+0.45}_{-0.46}$  \\
0.6&200&$0.748^{+0.014}_{-0.011}$   & $0.726^{+0.014}_{-0.007}$  & $4.36^{+0.56}_{-0.42}$  \\
0.4&300& $0.505^{+0.026}_{-0.009}$  & $0.494^{+0.008}_{-0.008}$  & $3.31^{+0.40}_{-0.37}$  \\
0.4&2000& $0.512^{+0.011}_{-0.010}$ & $0.503^{+0.001}_{-0.007}$   & $3.35^{+0.37}_{-0.38}$ \\
\end{tabular}
\caption{Cross-sections, subject to the cuts outlined in the text, for $Wt$ and $t\bar{t}$ production, obtained using MC@NLO. The single top results are obtained using both 
diagram removal (DR) and diagram subtraction (DS), and correspond to both top or antitop quarks in the final state. Quoted errors are due to scale variation by a factor of two.}
\label{DRvsDS}
\end{center}
\end{table}
The cross-sections have been obtained for strictly $Wt$ production, and then multiplied by a factor of two to account for $\bar{t}$ production. The uncertainties quoted correspond to varying the common renormalization and factorization scale in the range $m_t/2<\mu<2m_t$. From the table, one may note the following:
\begin{itemize}
\item The DR and DS results agree to within around 3\% in all cases, which is similar to the uncertainty in each result due to scale variation\footnote{Slightly more scale variation is observed if the factorization and renormalization scales are varied independently from each other. We checked that this does not invalidate the fact that the $Wt$ cross-section is larger than the scale variation uncertainty of the top pair production result, when $\mu_F$ and $\mu_R$ are varied such that their ratio is never more than $2$.}. Thus, the interference term between $Wt$ and $t\bar{t}$ production indeed appears to be small.
\item The $Wt$ cross-section is larger than the uncertainty on the $t\bar{t}$ cross-section due to scale variation. Thus, the $Wt$ signal is well-defined and visible above the $t\bar{t}$ background. 
\end{itemize}
As stressed above, both of these properties are needed before one can sensibly claim to be able to isolate $Wt$ production. Also, they are dependent on the cuts applied, and the above cuts are a fairly minimal choice such that both of these requirements can be satisfied.\\

\begin{figure}[h]
\begin{center}
\begin{tabular}{cc}
\scalebox{0.38}{\includegraphics{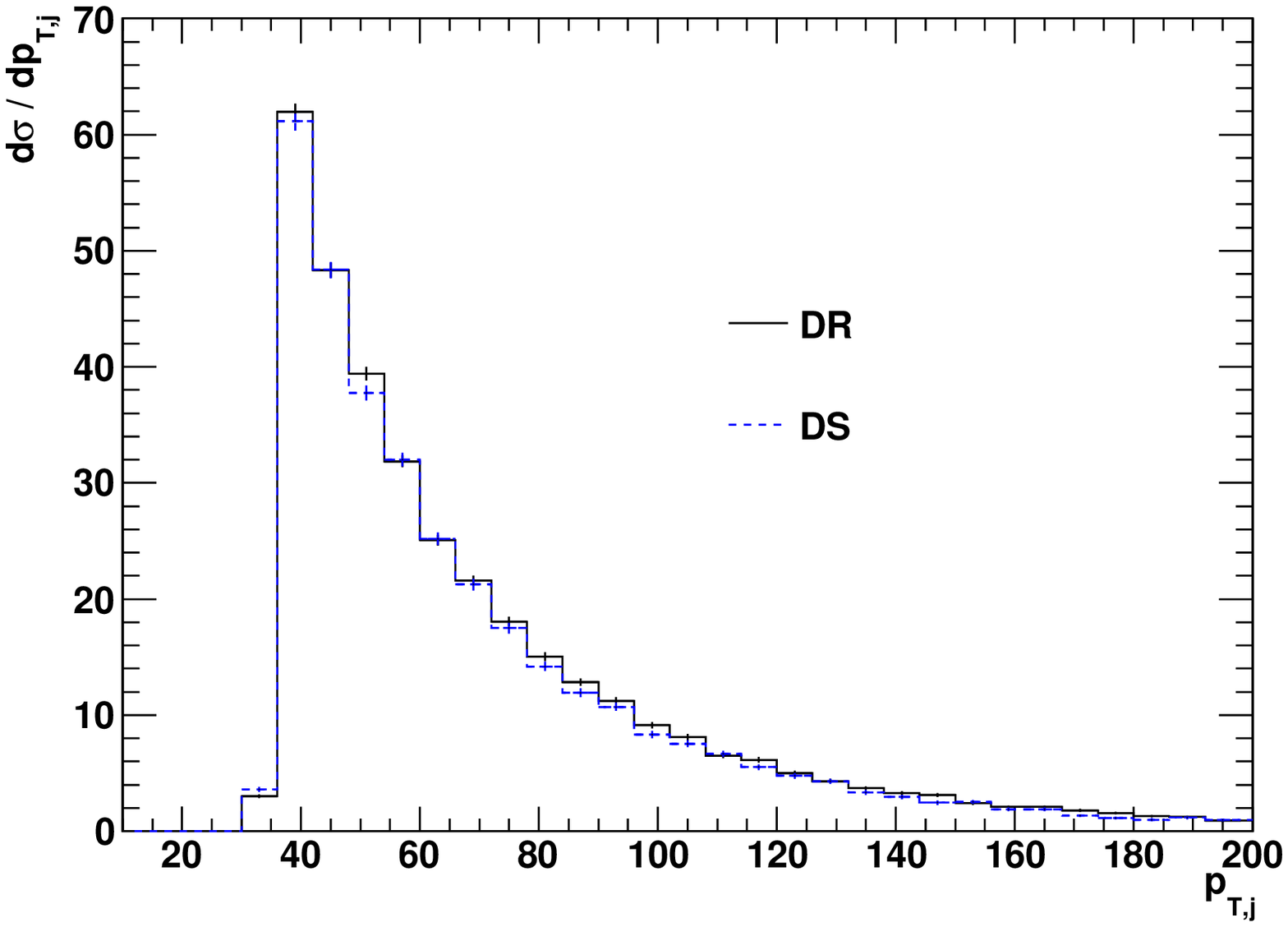}}&\scalebox{0.38}{\includegraphics{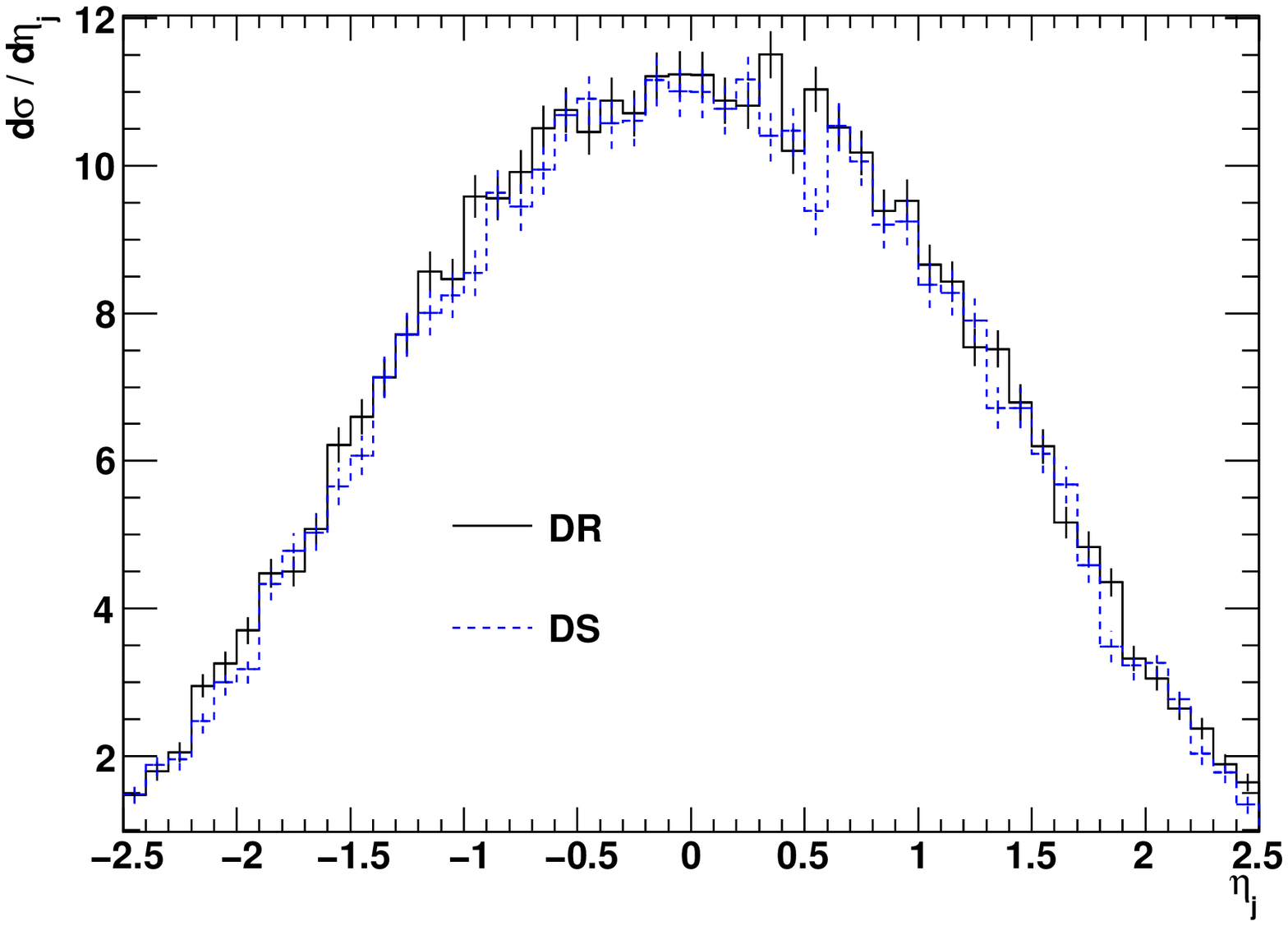}}\\
(a) & (b)
\end{tabular}
\caption{The transverse momentum (a) and pseudo-rapidity (b) distributions of the light jets in $Wt$ production (subject to the cuts outlined in the text), shown for both diagram removal (DR) and diagram subtraction (DS). The $b$-tagging efficiency and light jet rejection rate are given by $e_b=0.6$ and $r_{lj}$=30 respectively. Uncertainties are statistical, and the vertical axis has arbitrary normalization.}
\label{ptjDRvsDS}
\end{center}
\end{figure}

\begin{figure}[h]
\begin{center}
\begin{tabular}{cc}
\scalebox{0.38}{\includegraphics{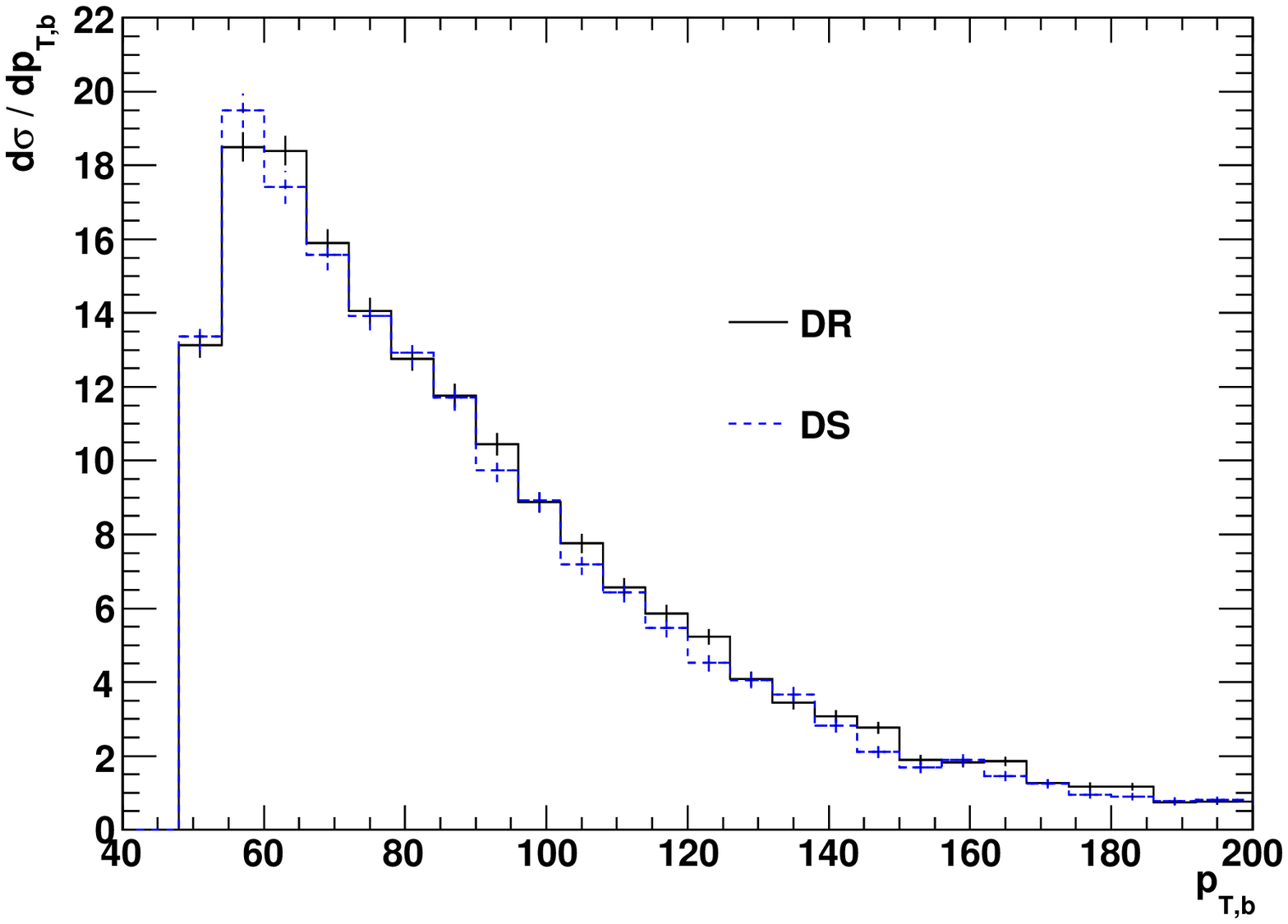}}&  \scalebox{0.38}{\includegraphics{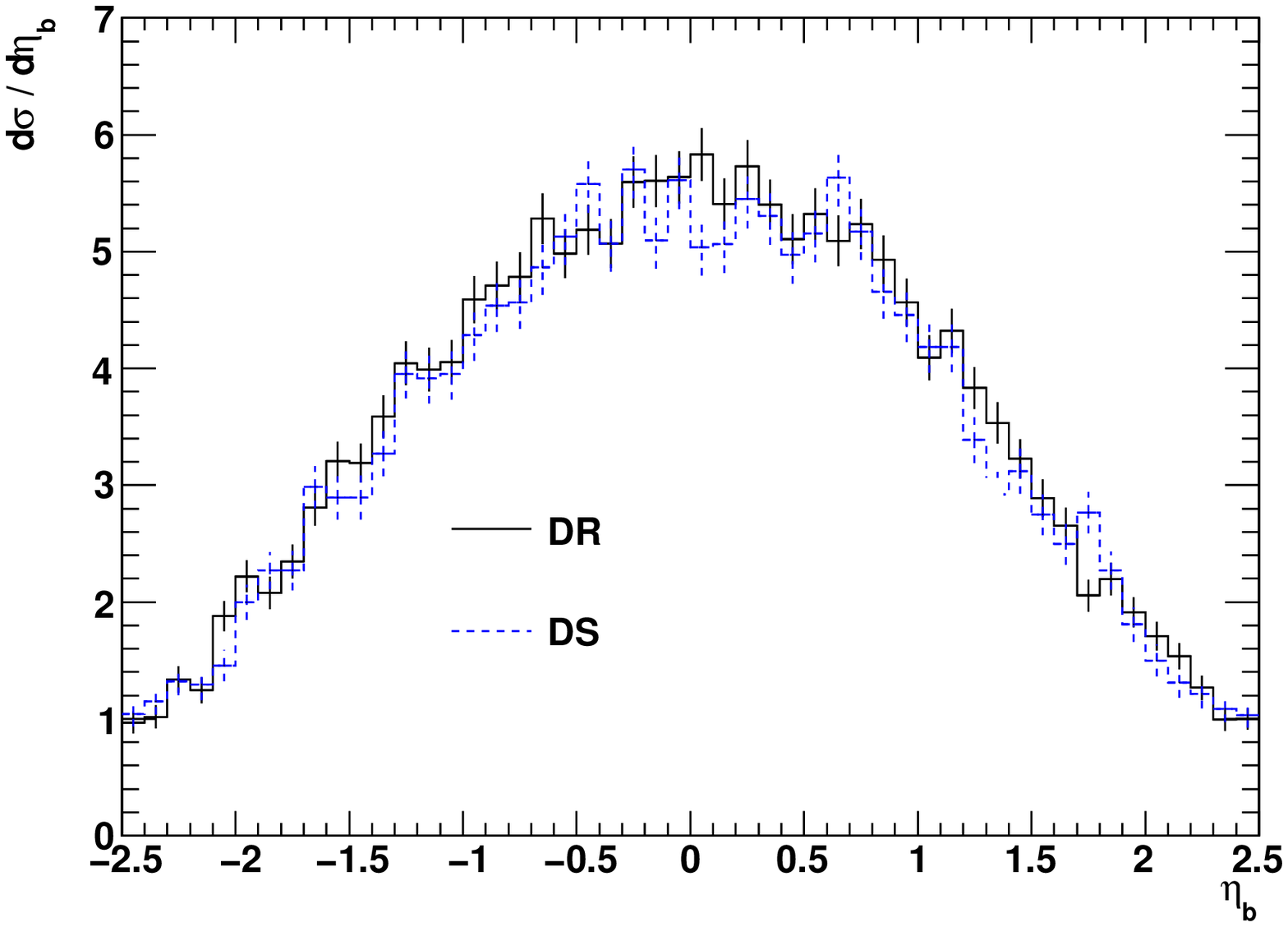}} \\
(a) & (b)
\end{tabular}
\caption{The transverse momentum (a) and pseudo-rapidity (b) distributions of the hard $b$ jet in $Wt$ production (subject to the cuts outlined in the text), shown for both diagram removal (DR) and diagram subtraction (DS). The $b$-tagging efficiency and light jet rejection rate are given by $e_b=0.6$ and $r_{lj}$=30 respectively. Uncertainties are statistical, and the vertical axis has arbitrary normalization.}
\label{pbjDRvsDS}
\end{center}
\end{figure}

\begin{figure}[h]
\begin{center}
\begin{tabular}{cc}
\scalebox{0.38}{\includegraphics{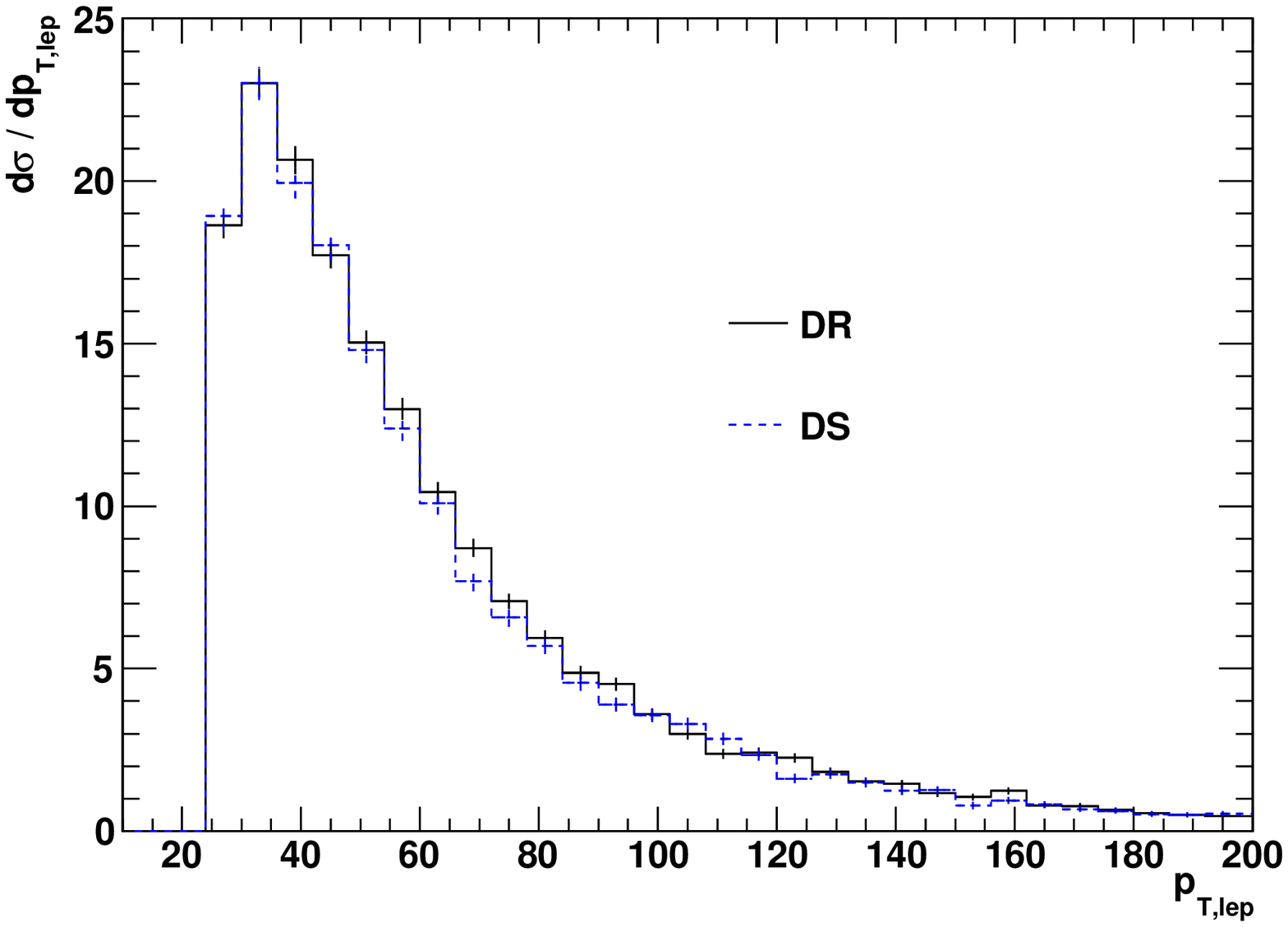}}&\scalebox{0.38}{\includegraphics{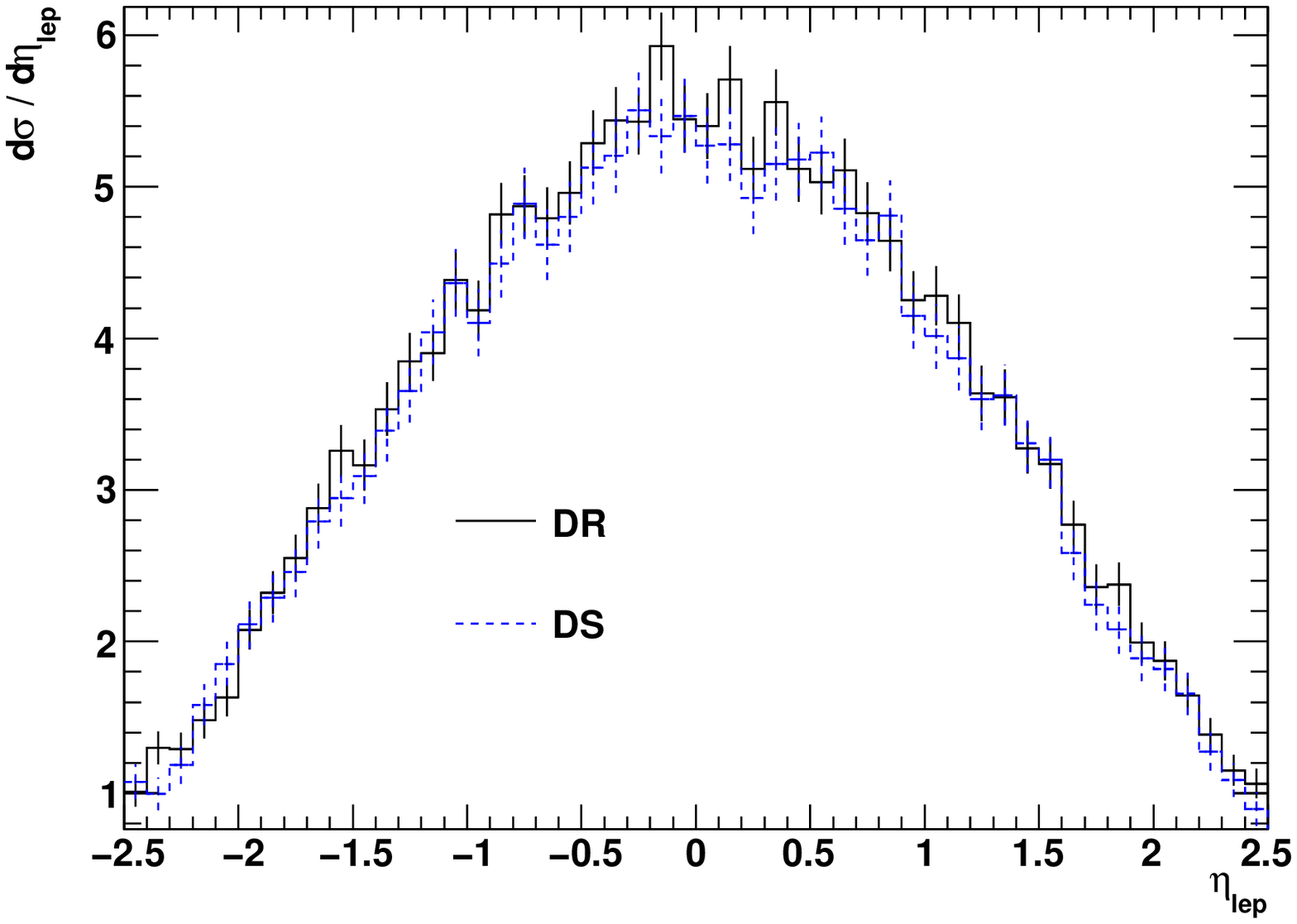}}\\
(a) & (b)
\end{tabular}
\caption{The transverse momentum (a) and pseudo-rapidity (b) distributions of the isolated lepton in $Wt$ production (subject to the cuts outlined in the text), shown for both diagram removal (DR) and diagram subtraction (DS). The $b$-tagging efficiency and light jet rejection rate are given by $e_b=0.6$ and $r_{lj}$=30 respectively. Uncertainties are statistical, and the vertical axis has arbitrary normalization.}
\label{ptlDRvsDS}
\end{center}
\end{figure}

Although DR and DS agree at the total cross-section level, it is also important to verify the agreement in kinematic distributions. This is possible given that both DR and DS are defined in a parton shower context at the fully exclusive level i.e. locally in phase space. As examples, in figures \ref{ptjDRvsDS}-\ref{ptlDRvsDS} we show the transverse momentum and rapidity distributions of the light jets, $b$ jet and isolated lepton entering the cuts defined above. One sees that agreement is obtained within statistical uncertainties, in addition to the agreement within scale uncertainties noted above. \\

\begin{figure}[h]
\begin{center}
\begin{tabular}{cc}
\scalebox{0.38}{\includegraphics{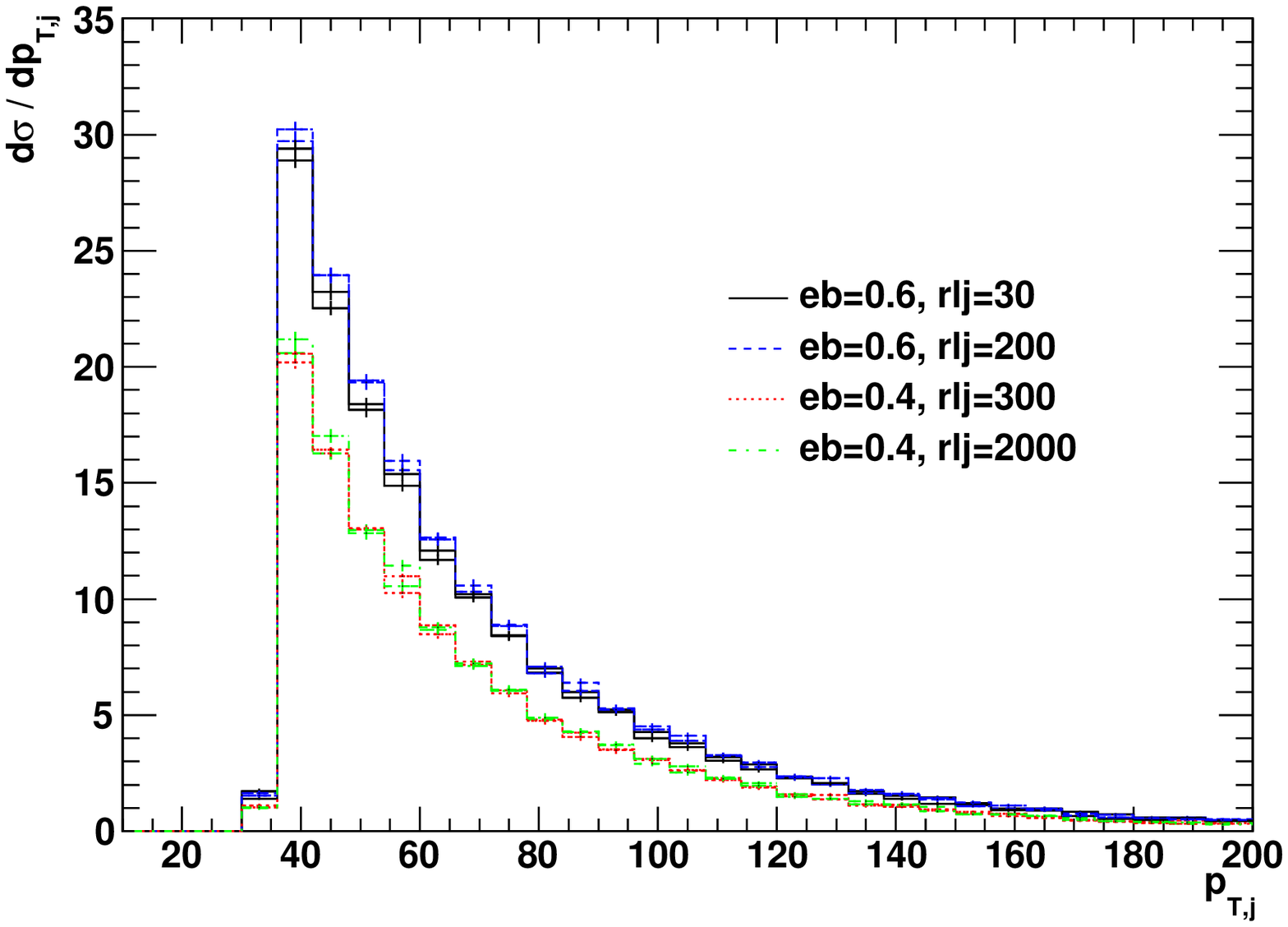}}& \scalebox{0.38}{\includegraphics{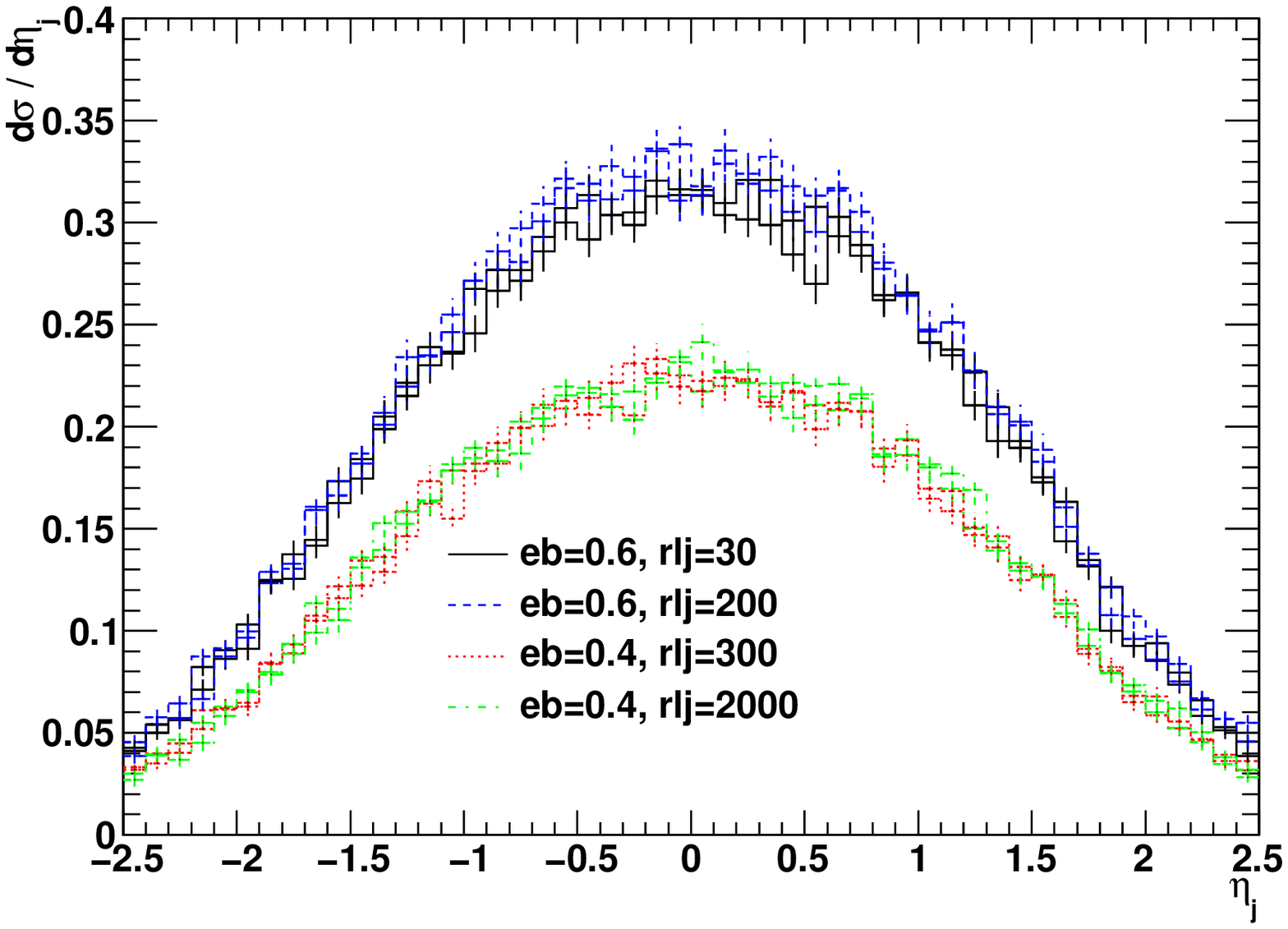}} \\
(a) & (b)
\end{tabular}
\caption{The transverse momentum (a) and pseudo-rapidity (b) distributions of the light jets in $Wt$ production, shown for various choices of $b$-tagging efficiency $e_b$ and light jet rejection rate $r_{lj}$ (normalized to the first choice). Results are shown for both diagram removal (DR) and diagram subtraction (DS).}
\label{ptj_ebrj}
\end{center}
\end{figure}

\begin{figure}[h]
\begin{center}
\begin{tabular}{cc}
\scalebox{0.38}{\includegraphics{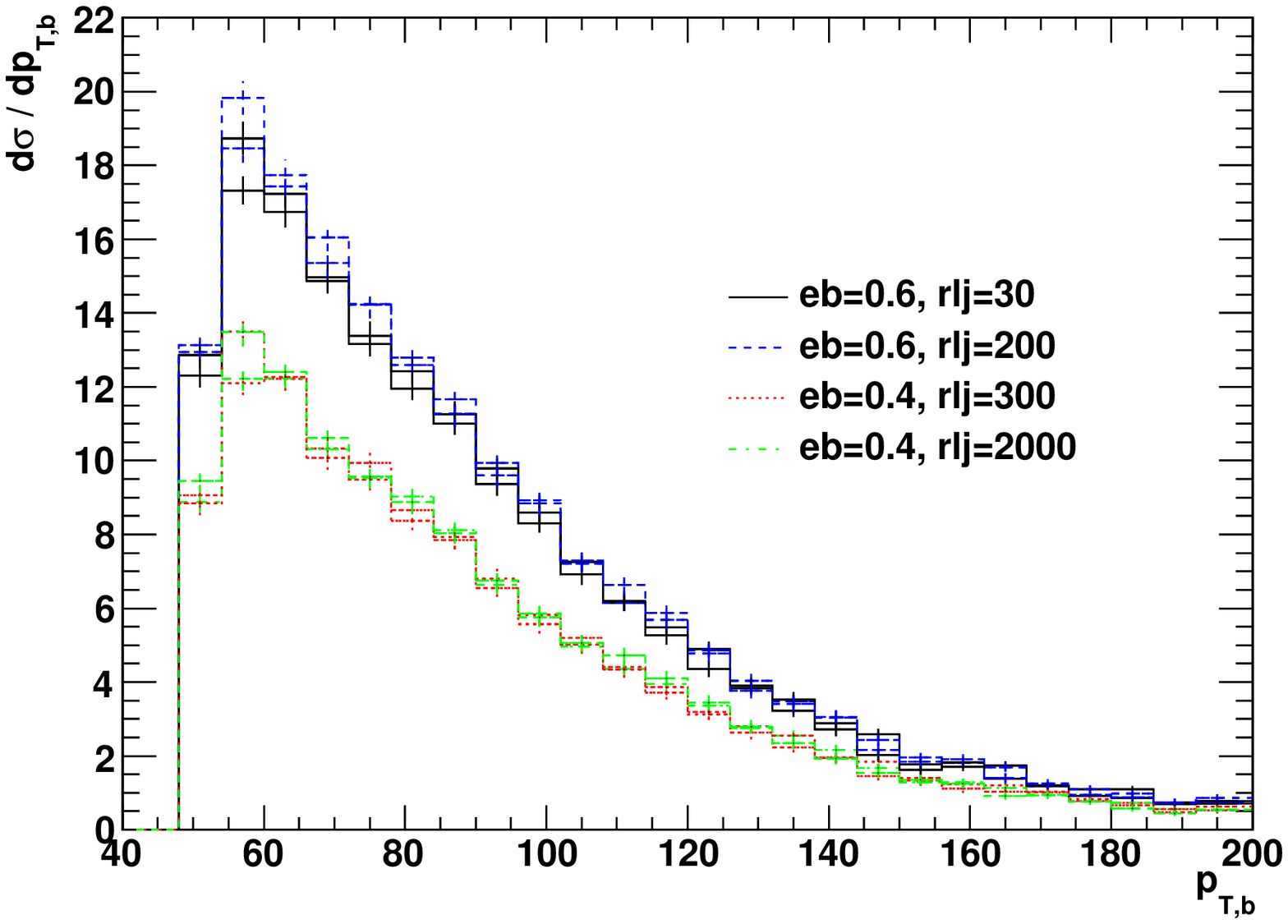}}& \scalebox{0.38}{\includegraphics{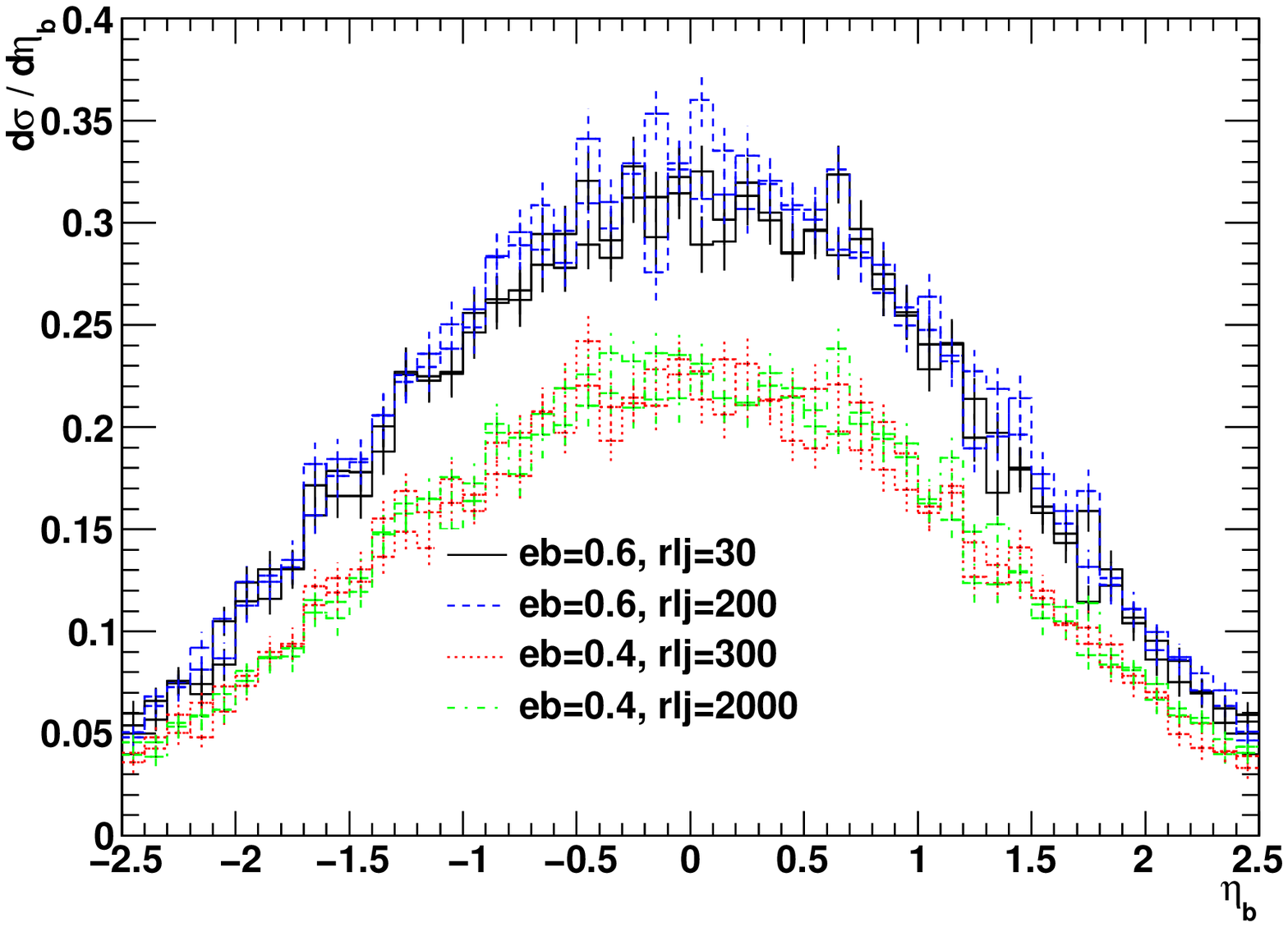}}\\
(a) & (b)
\end{tabular}
\caption{The transverse momentum (a) and pseudo-rapidity (b) distributions of the $b$ jet in $Wt$ production, shown for various choices of $b$-tagging efficiency $e_b$ and light jet rejection rate $r_{lj}$ (normalized to the first choice). Results are shown for both diagram removal (DR) and diagram subtraction (DS).}
\label{etab_ebrj}
\end{center}
\end{figure}

One must also consider distributions for various choices of $b$ tagging efficiency and light jet rejection rate. Of these, the former has a potentially damaging effect on the ability of jet cuts to reduce the $Wt$-$t\bar{t}$ interference, as these rely on cutting out events with a second hard $b$ jet. The transverse momentum and rapidity distributions for the light and $b$ jets are shown, for all four non-trivial choices of $e_b$ and $r_{lj}$ given in table \ref{DRvsDS}, in figures \ref{ptj_ebrj}-\ref{etab_ebrj}.\\

One sees good agreement between the DR and DS results for all choices of $e_b$ and $r_{lj}$ and all distributions. Thus, the above cuts do isolate $Wt$ production in a well-defined sense. Note that the ratio of the $Wt$ and $t\bar{t}$ cross-sections is $\simeq1:4.7$ (before accounting for $b$-tagging efficiency and light jet rejection). The above, however, is a rough analysis designed to address interference issues. Additional observables can be used to further enhance the signal without diminishing the cross-section too much (see e.g.~\cite{Aad:2009wy}). However, it is encouraging that even without a highly optimized signal to background ratio, the $Wt$ signal is well-defined.\\

\begin{figure}[h]
\begin{center}
\begin{tabular}{cc}
\scalebox{0.38}{\includegraphics{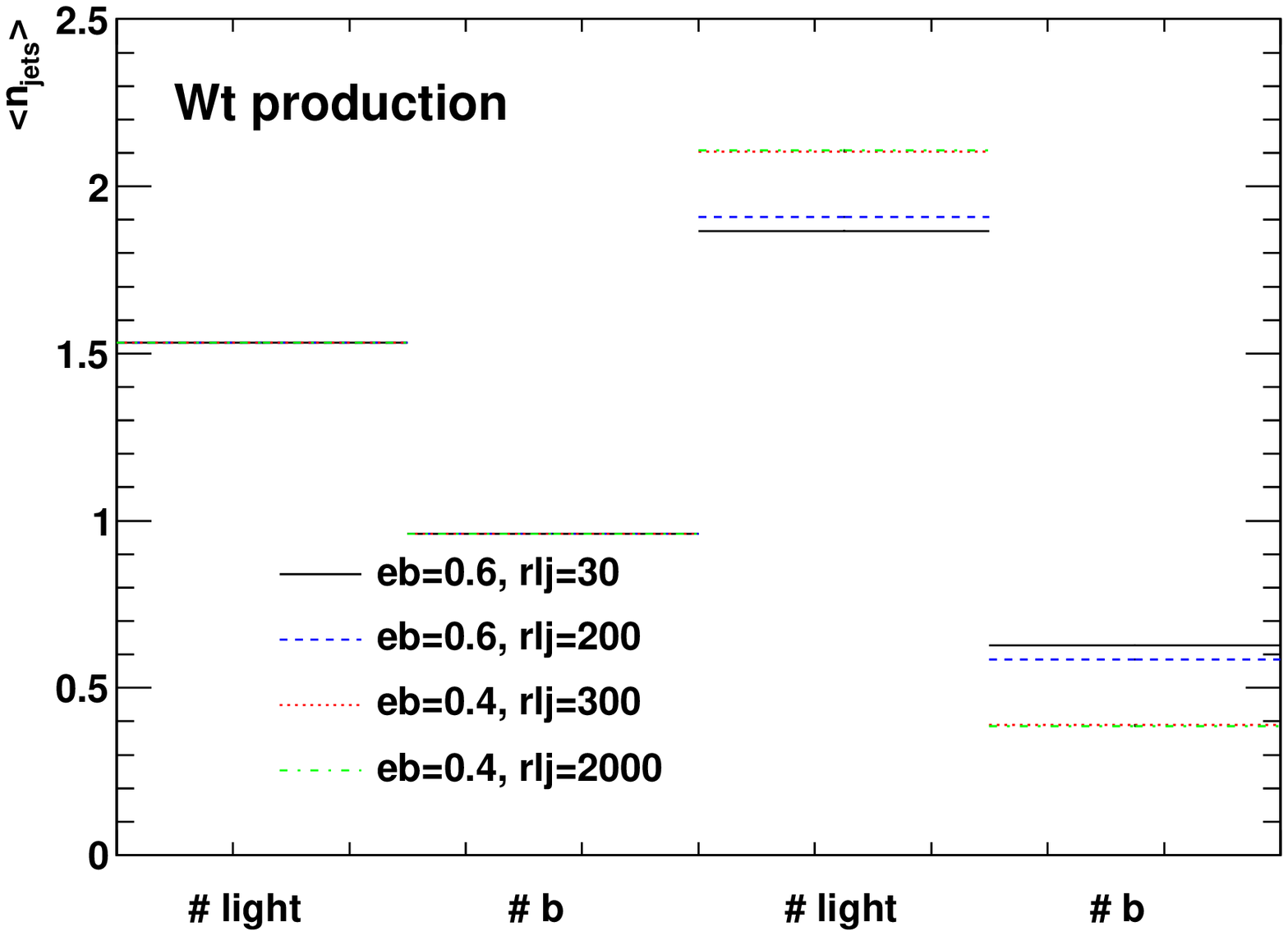}}& \scalebox{0.38}{\includegraphics{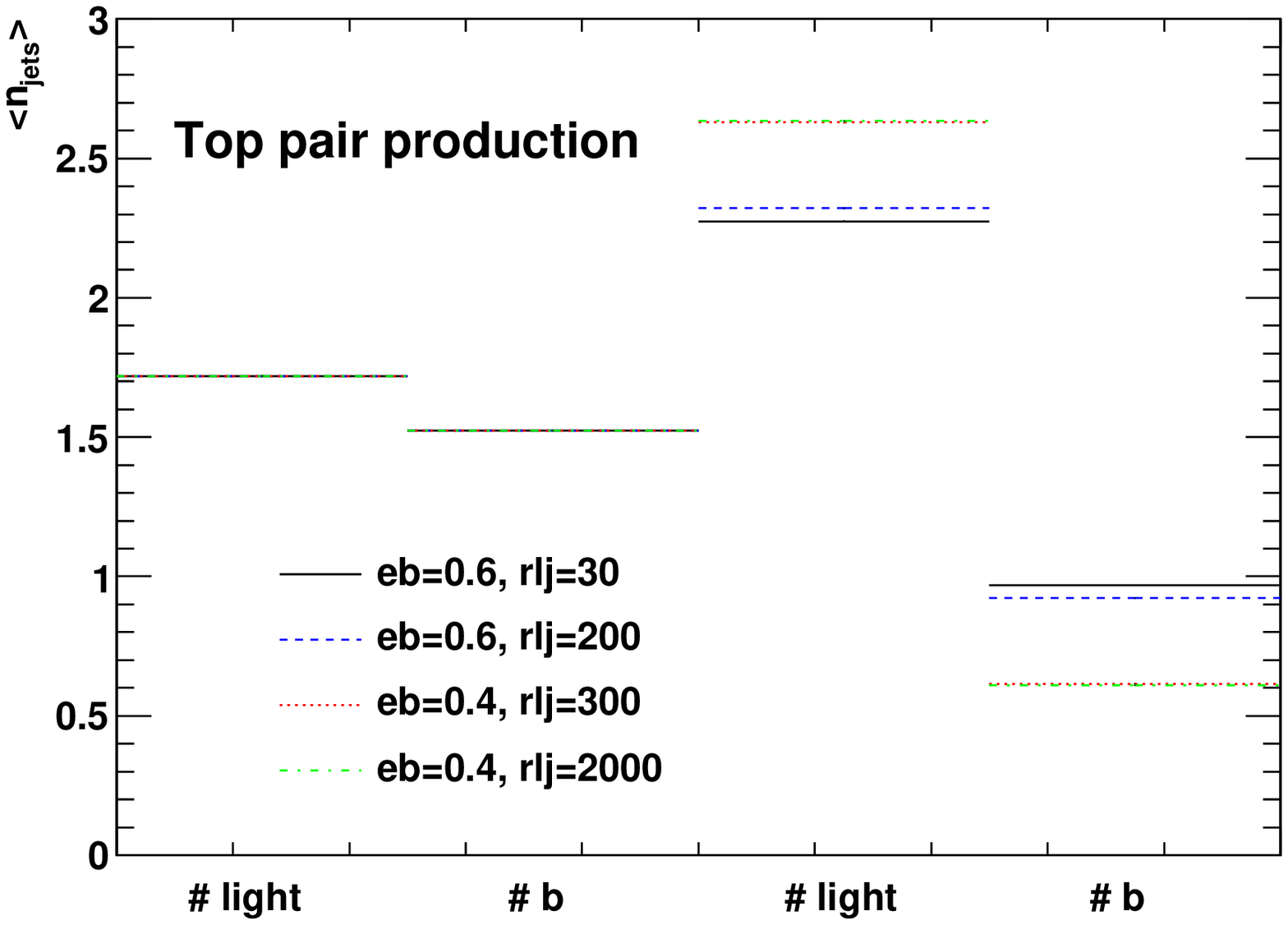}}\\
(a) & (b)
\end{tabular}
\caption{The average number of light and $b$ jets before (left-hand bins) and after (right-hand bins) reshuffling due to $b$-tagging efficiency and light-jet rejection rate. The $Wt$ results have been obtained using diagram removal (DR).}
\label{avgjets}
\end{center}
\end{figure}

The effect of $b$ tagging efficiency and light-jet rejection rate can be further appreciated by looking at figure~\ref{avgjets}, which shows the average number of $b$ and light jets per event (satisfying the detector cuts $p_T>25$ GeV and $|\eta|<2.5$, but before the full $Wt$ signal cuts have been applied) before and after reshuffling due to non-trivial $e_b$ and $r_{lj}$. One sees that the average number of $b$ jets is slightly below one for $Wt$ production, even before reshuffling. Given that a hard $b$ jet is required by the signal cuts, this makes the $Wt$ cross-section more sensitive to $b$ tagging efficiency than that of top pair production, as can be seen directly in table~\ref{DRvsDS}.\\

\begin{figure}
\begin{center}
\scalebox{0.38}{\includegraphics{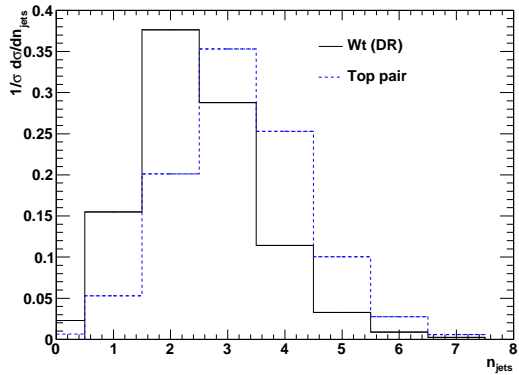}}
\caption{Distribution in the total number of jets which pass the detector cuts $p_T>25$ GeV and $|\eta|<2.5$.}
\label{njets}
\end{center}
\end{figure}

In figure~\ref{njets} we show the total number of jets (light plus $b$ jets) passing the detector cuts. One clearly sees that top pair production has higher jet multiplicities on average, hence the efficacy of the signal cuts in selecting $Wt$ production. Furthermore, there is a non-trivial fraction of events with five or more hard jets. This, combined with the fact that the signal cuts require three jets, suggests that a parton shower framework (rather than a fixed order matrix element) is indeed more appropriate for describing $Wt$ production, given the limited number of partons in presently available fixed order matrix element calculations. There is another reason why a parton shower framework is more appropriate, namely that one does not necessarily trust a fixed order matrix element description of emitted partons at lower transverse momenta, such as those ($\simeq 25$ GeV) involved in the jet veto cuts (see section 5.1 of~\cite{Frixione:2008yi} for a discussion related to this point).\\

A comment is in order regarding the use of a sequential cut method in order to isolate the $Wt$ signal, when recent experimental analyses rely more heavily on methods based on neural networks, boosted decision trees (BDT) and matrix element methods (e.g. the recent discovery of single top production at the Tevatron~\cite{Aaltonen:2009jj,Abazov:2009ii}). It is very likely that such methods will be applied at the LHC in order to isolate $Wt$ production. For example, a sample analysis (at LO plus parton shower level) is presented by the ATLAS experiment in~\cite{Aad:2009wy}, alongside a traditional sequential cut analysis. It is not always clear how systematic uncertainties in Monte Carlo models propagate through such analyses, including in this case the uncertainty attached with separating $Wt$ and $t\bar{t}$. The safest way to proceed, in cases where there is any doubt, is to repeat a given analysis which depends on the use of MC@NLO for $Wt$ production using both the DR and DS options.
\section{$Wt$ production as a background to $H\rightarrow WW$}
\label{Higgs}
In the previous section, we have shown that it is possible to isolate $Wt$ production as a signal. However, this is not the only context in which $Wt$ production occurs - one must also consider it as a background to other production processes. In such cases (and as suggested by the results of the previous section), one wishes to use as accurate a description of the background as possible, which strongly motivates the use of MC@NLO. However, one must check in such a case that this description is well-defined, namely that DR and DS agree for the cuts used to isolate the signal of interest. If this turns out to be true, one may reliably estimate the top background to the production process of interest by combining samples of $t\bar{t}$ and $Wt$ events (corresponding to an incoherent sum of the hard processes).\\

In this section, we consider an example of $Wt$ and $t\bar{t}$ as backgrounds to a third process, that of Higgs boson production with subsequent decay to a pair of $W$ bosons. This is of topical interest, given that the $H\rightarrow W^+W^-$ decay mode dominates for intermediate Higgs boson masses $150$ GeV $\lesssim m_H\lesssim180$ GeV, making this the only viable discovery channel in this window. Furthermore, the dominant background is from top pair production (with single top processes also significant), thus this is an excellent example to illustrate the use of $Wt$ production as a background. Our aim here is not to present a detailed phenomenological study of Higgs boson production (see p.110 of~\cite{Aad:2009wy} for an up-to-date experimental study), but rather to examine whether MC@NLO can be used to reliably estimate the $Wt$ background.\\

In order to minimize QCD jet backgrounds, it is common to consider the case where both $W$ bosons stemming from the Higgs boson decay leptonically i.e.
\begin{equation}
H\rightarrow W^+W^-\rightarrow l_1^+\bar{\nu}_{1}l_2^-\nu_{2},
\label{HWWll}
\end{equation}
where $l_i$ is either an electron or muon, and $\nu_{i}$ its corresponding neutrino. Then spin correlations can be used to efficiently isolate the signal against top-related backgrounds~\cite{Dittmar:1996ss} (see also~\cite{Davatz:2006kb,Davatz:2006ja,Davatz:2004zg,Davatz:2006ut,Anastasiou:2007mz,Anastasiou:2008ik}). Motivated by~\cite{Anastasiou:2007mz,Anastasiou:2008ik,Kauer:2000hi}, we use the following example cuts to isolate the Higgs signal:
\begin{center}
\textbf{Higgs signal cuts}
\end{center}
\begin{enumerate}
\item There must be two opposite sign leptons satisfying $p_T>25$ GeV and $|\eta|<2.5$.
\item The invariant mass of the charged lepton pair should satisfy $12$ GeV$<m_{ll}<40$ GeV.
\item The azimuthal angle between the leptons (i.e. the angle in the transverse plane) should be less than $\pi/4$.
\item The lepton with the highest $p_T$ should satisfy $30$ GeV$<p_T<$55 GeV.
\item There must be a missing transverse energy of at least $50$ GeV.
\item There must be no jets (i.e. either $b$ or light jets) with $p_T>25$ GeV and $|\eta|<2.5$.
\end{enumerate}
More sophisticated cuts require isolation of the leptons from hadronic activity, as well as tuning of the various parameters introduced above. However, as in section~\ref{signal}, we choose a reasonably minimal set of cuts associated with the signal of interest. Conclusions reached about whether the $Wt$ background can be well-defined will then apply in more realistic analyses.\\

Of the above cuts, the jet veto (i.e. cut number 6) is particularly effective in reducing the background from top quark production, either singly or in pairs. One could again consider various $b$ tagging efficiencies $e_b$ and light jet rejection rates $r_{lj}$, but given that the jet veto applies to the total number of jets, these will be irrelevant in our analysis. In the results that follow we use parton densities, as well as top and $W$ masses and widths, as described in section~\ref{signal}. Our default factorization and renormalization scale choices are again $\mu_R=\mu_F=m_t$, and we allow electrons or muons in the decay of the $W$ bosons.\\

For the above choice of signal cuts, the Higgs signal cross-section is (using MC@NLO with a renormalization and factorization scale equal to the Higgs mass) 81.8 fb for a Higgs boson mass $m_H=165$ GeV. This is comparable to the corresponding figure presented in~\cite{Anastasiou:2007mz,Anastasiou:2008ik}, although slightly higher due to the requirement in that paper that the leptons be isolated from hadronic activity\footnote{To obtain the above number one must include spin correlations in the decay of the Higgs boson, particularly given the cut on the azimuthal angle between the lepton pair. These are not implemented in the latest public release of HERWIG, hence we use the unreleased version referred to in~\cite{Anastasiou:2007mz,Anastasiou:2008ik}.}. After cuts, the backgrounds due to top quark production are somewhat smaller than the background from non-resonant $W$ pair production~\cite{Dittmar:1996ss}, but are still significant. Our results for the top pair and $Wt$ backgrounds are shown in table~\ref{higgsresults}. \\
\begin{table}
\begin{center}
\begin{tabular}{c|c}
Process&$\sigma_{NLO}$/fb\\
\hline
$h\rightarrow WW$&81.8 $\pm$0.4 \\
$t\bar{t}$&12.25 $\pm$ 0.3\\
$Wt$ (DR) & 6.91 $\pm$ 0.06\\
$Wt$ (DS) & 6.89 $\pm$ 0.07
\end{tabular}
\caption{Cross-sections obtained using MC@NLO for the $H\rightarrow W^+W^-$ signal cuts described in the text, where the $W$ bosons can decay to electrons or muons. Note that the $Wt$ results include both top or antitop quarks in the final state. Uncertainties correspond to statistical errors only.}
\label{higgsresults}
\end{center}
\end{table}

\begin{figure}[h]
\begin{center}
\begin{tabular}{cc}
\scalebox{0.38}{\includegraphics{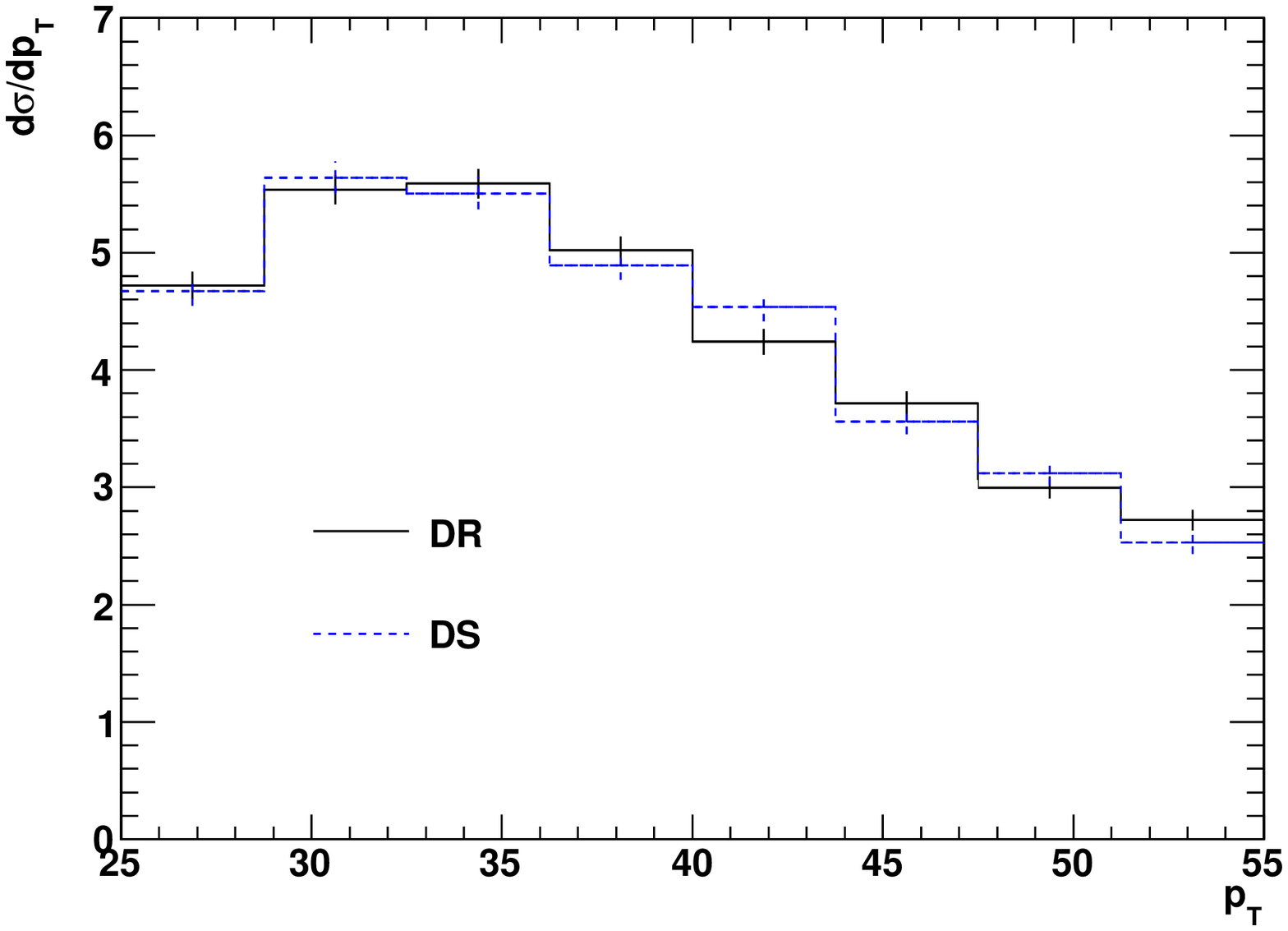}} &\scalebox{0.38}{\includegraphics{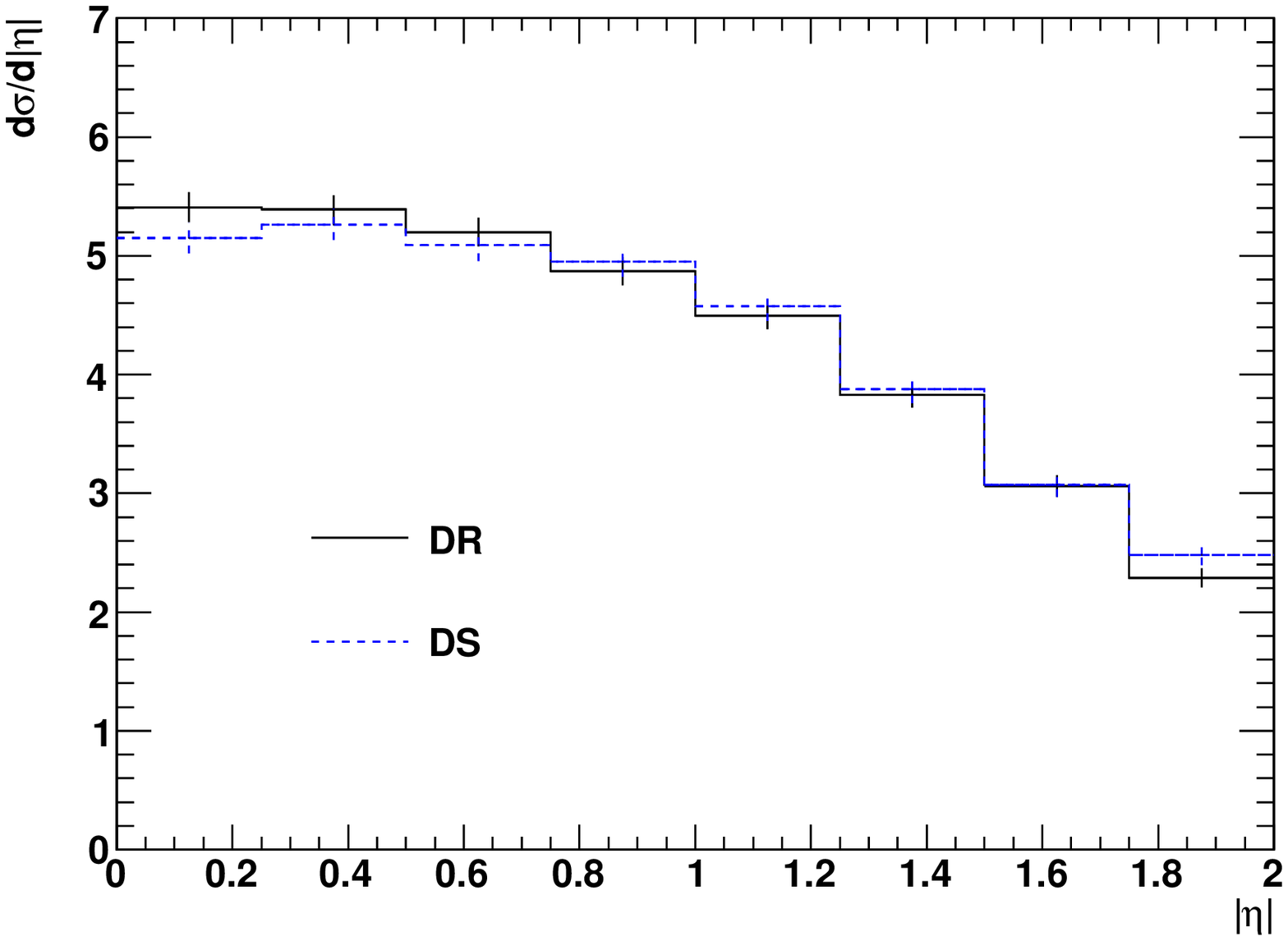}}  \\
(a)&(b)
\end{tabular}
\caption{The transverse momentum (a) and absolute pseudo-rapidity (b) distributions of the lepton from the top quark in $Wt$ production subject to the Higgs signal cuts described in the text, obtained using DR (black) and DS (blue). Uncertainties (indicated by the vertical bars) are statistical, and the vertical axis shows arbitrary units.}
\label{pt1}
\end{center}
\end{figure}

\begin{figure}[h]
\begin{center}
\begin{tabular}{cc}
\scalebox{0.38}{\includegraphics{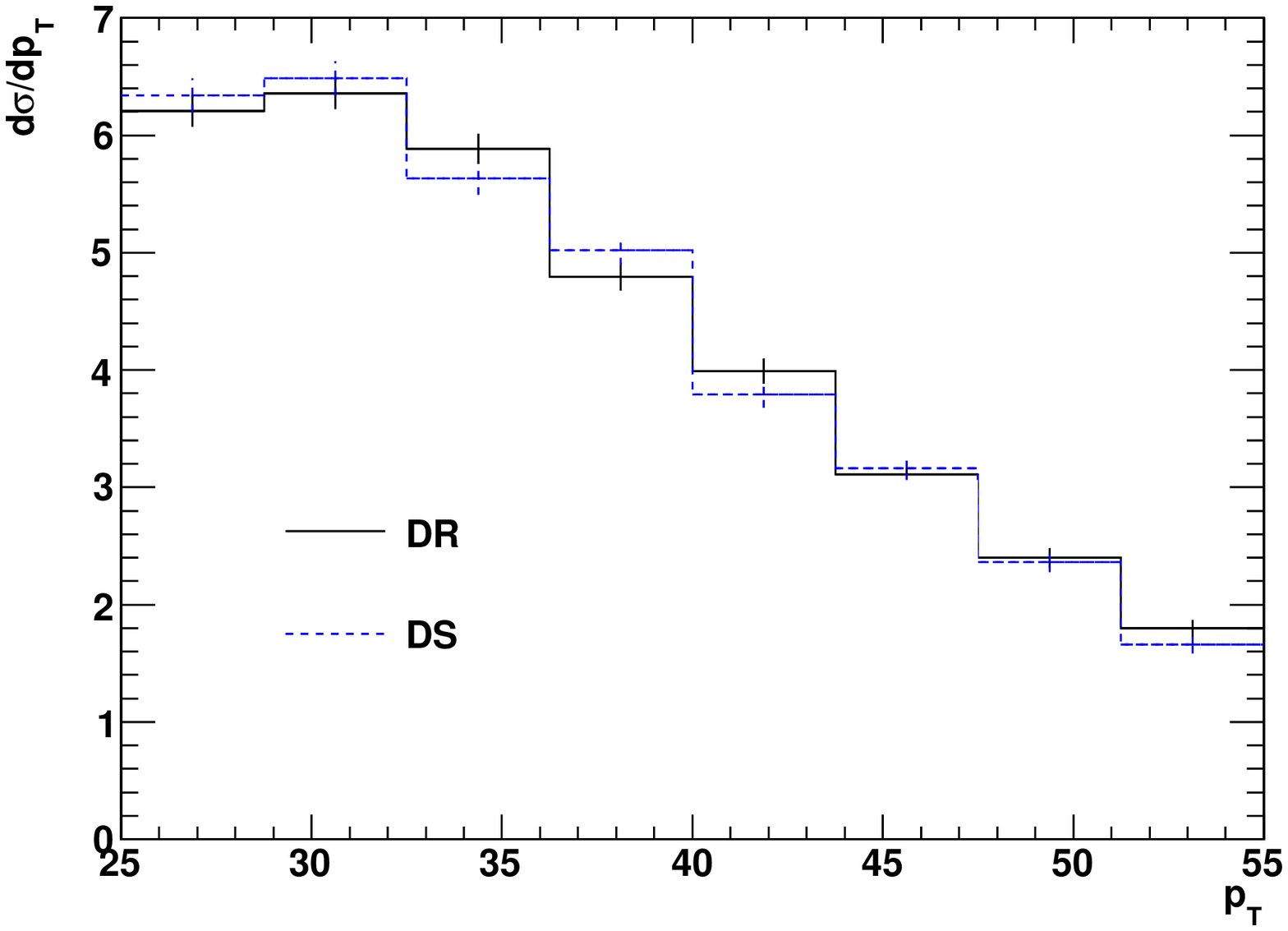}}&\scalebox{0.38}{\includegraphics{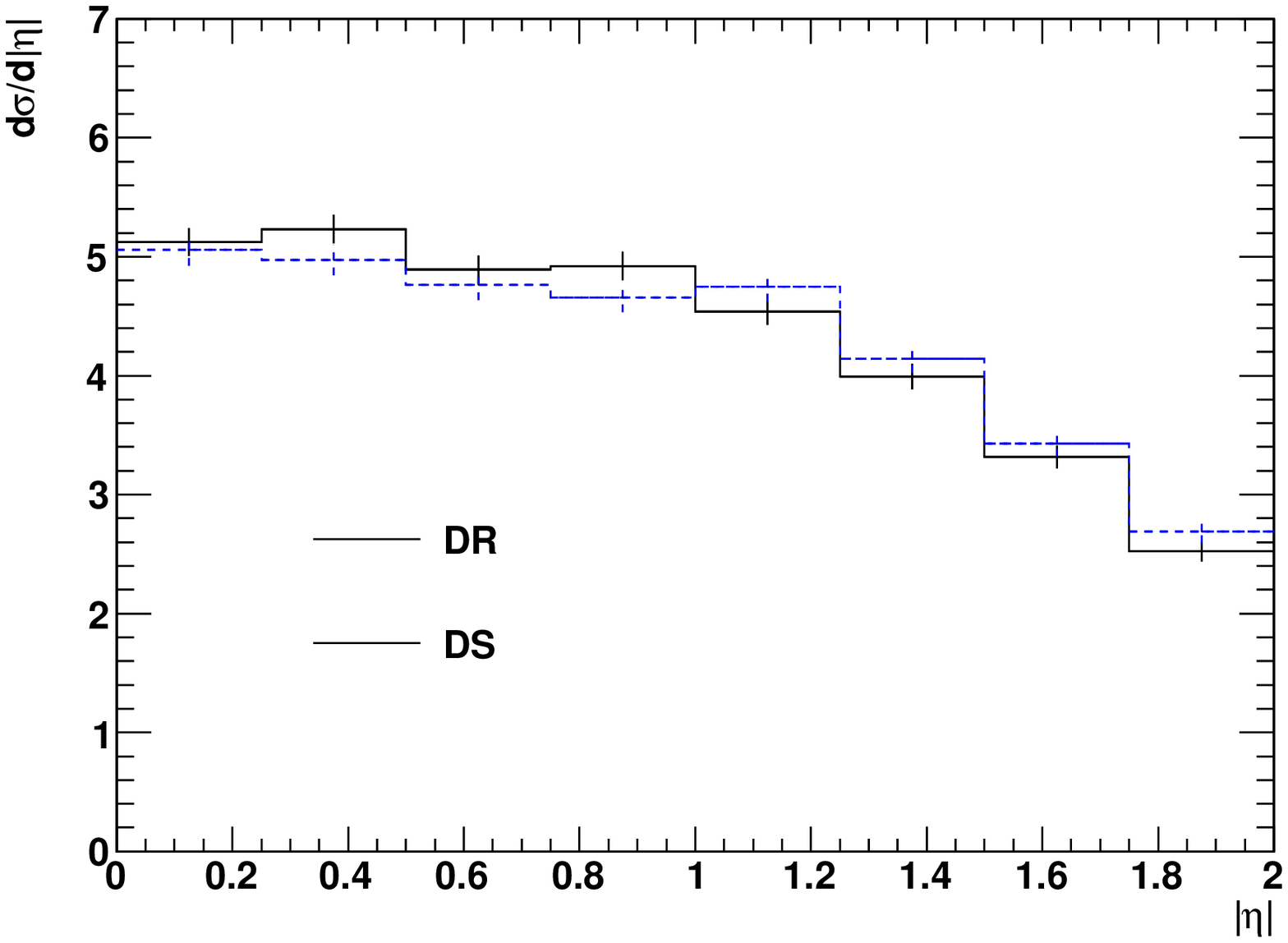}}\\
(a)&(b)
\end{tabular}
\caption{The transverse momentum (a) and absolute pseudo-rapidity (b) distributions of the lepton from the $W$ boson in $Wt$ production subject to the Higgs signal cuts described in the text, obtained using DR (black) and DS (blue). Uncertainties are statistical, and the vertical axis shows arbitrary units.}
\label{eta2}
\end{center}
\end{figure}

One sees that the $Wt$ background is more than half the size of the top pair background. That these are similar in magnitude is not surprising, given the jet veto involved in the selection cuts. Importantly, the DR and DS results for $Wt$ production agree well within statistical uncertainties (we checked that these are larger in this case than the uncertainty that results from varying the common renormalization and factorization scale by a factor of two). As in section~\ref{signal}, it is important to check that kinematic distributions also agree well when calculated with both DR and DS. Some examples are shown in figures~\ref{pt1}-\ref{eta2}, namely the transverse and absolute pseudo-rapidity distributions of the two final state leptons. One sees that the DR and DS results agree closely within statistical uncertainties. \\

We have seen so far that when top production occurs as a background to a given process (namely Higgs boson production with subsequent decay to $W$ bosons), one is still able to define $Wt$ production as a separate background subject to the cuts used to isolate the signal. This means that in evaluating the combined background from top production, $t\bar{t}$ production and $Wt$ events can be generated separately, and the results added together without having to worry about interference effects.\\

Some remarks are in order regarding how many of the above statements can be generalized to other processes to which top production is a significant background. There are a number of possibilities in general:
\begin{itemize}
\item Top pair and $Wt$ production are comparable in cross-section, and a significant fraction of the signal cross-section, but such that the interference between $Wt$ and $t\bar{t}$ production is small. This is the case considered above.
\item Top pair and $Wt$ production are comparable in cross-section, and a significant fraction of the signal cross-section, such that the interference is not small. We discuss this case in more detail below.
\item Top pair and $Wt$ production are comparable in size, and their sum is an insignificant fraction of the signal. One does not have to worry about interference in this case, given that top pair production itself is not a significant background. 
\item Top pair production is a significant background, but $Wt$ production has a much lower cross-section. In general in this scenario interference between $Wt$ and $t\bar{t}$ is non-negligible, but owing to the small size of the $Wt$ cross-section is irrelevant. We will see an example of this in the following section, when $t\bar{t}$ itself is considered as the signal.
\end{itemize}
As is clear from the above categorization, one need only worry if the second situation occurs. This naturally presents two options. Either one can find an alternative to separating $Wt$ and $t\bar{t}$ production in order to estimate the background, or one can take the difference between DR and DS as a measure of systematic uncertainty. If this latter uncertainty is large, one concludes that it does not make sense to think of $Wt$ and $t\bar{t}$ as separate backgrounds. However, it seems likely that this latter situation only occurs in a minority of cases, given that most of the time one is trying to reduce both $Wt$ and $t\bar{t}$ production as backgrounds. Given the $t\bar{t}$ cross-section is generically larger than the $Wt$ cross-section, any successful reduction of the top pair background will usually render the $Wt$ interference insignificant. \\

Ultimately, one expects the MC@NLO calculation for the sum of $Wt$ and $t\bar{t}$ production to be a good approximation in many cases. One may worry in cases where top backgrounds remain large, and the signal cuts do not decrease the ratio of top pair to single top production. If in doubt, one may run the DR and DS codes separately, and thus quantify the systematic uncertainty due to interference effects. Whether or not this uncertainty is significant depends on the process, and also on the other systematic uncertainties (e.g. scale variation) involved.
\section{Comparison with $WWbb$}
\label{finalstates}
In the previous sections, we saw that one can indeed recover $Wt$ as a well-defined process at the LHC, when trying to isolate and measure its properties. We also found that this was the case when single and top pair production were considered as backgrounds to a third process, namely Higgs boson production with subsequent decay of the latter into a $W$ boson pair. The analysis in both cases relied upon two things. Firstly, that  one has a way of quantifying the effect of interference between $Wt$ and $t\bar{t}$ production (such as the DR and DS codes of MC@NLO). The systematic uncertainty due to interference can then be meaningfully compared with other uncertainties in the problem (such as that due to scale variation or statistical uncertainty of the DR or DS results), in order to determine whether the $Wt$ mode makes sense. Secondly, that this interference can be reduced through adequate cuts.\\

Nevertheless, $Wt$ production is not strictly well-defined over all of phase space. In regions were the invariant mass of possible $Wb$ pairs not coming from the primary top approaches the top mass, the difference between DR and DS is potentially large. It can thus be objected that it is questionable to try to consider $Wt$ and $t\bar{t}$ as separate scattering processes, and to only consider given final states (which {\it are} well-defined). We consider such an approach in this section.\\

In the calculational framework adopted in previous sections (i.e. in which initial state $b$ quarks are present), the final states relevant to the coherent sum of $Wt$ and $t\bar{t}$ production are $WWb$ and $WWbb$, as discussed in section~\ref{intro}. Our aim is to calculate the top quark contributions to these final states, and compare the results with the description of the sum of the $Wt$ and $t\bar{t}$ processes obtained in the previous sections. Thus, we do not consider other processes which contribute to these final states (such as non-resonant $W$ pair production). \\

In order to obtain reliable predictions, one must combine the $WWb$ and $WWbb$ final states, and preferably interface the output to a parton shower. This raises a number of technical challenges (for a detailed discussion in a similar context to this paper, see~\cite{Alwall:2008qv}). One must avoid the double-counting that results from the presence of initial state $b$ quarks, and diagrams in which $b$ quark pairs are produced by gluon splitting (see~\cite{Kersevan:2006fq} for a discussion in the context of Monte Carlo generators). Furthermore, one must apply a matching procedure (e.g. CKKW~\cite{Catani:2001cc} or MLM~\cite{Caravaglios:1998yr}) owing to the presence of NLO real corrections to the LO $Wt$ process (i.e. $WWbb$ corrections to $WWb$). How to do this using presently available tools is not clear, given that in semileptonic decays of the two $W$ bosons, not all of the final state partons are of QCD origin.\\

\begin{figure}[h]
\begin{center}
\scalebox{1.0}{\includegraphics{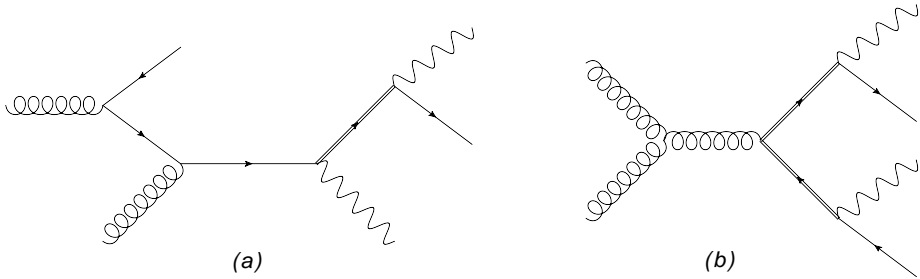}}
\caption{(a) Singly and (b) doubly resonant contributions to the $WWbb$ final state, where all $b$ quarks are explicitly produced via gluon splitting.}
\label{treediags}
\end{center}
\end{figure}

In order to circumvent these difficulties, we consider in this section a fixed flavor scheme in which the bottom quark parton density is not present. All initial state $b$ quarks entering the hard interaction are then explicitly generated from gluon splitting, as shown (for LO $Wt$ production) in figure~\ref{treediags}(a)\footnote{A similar calculation was considered in~\cite{Kauer:2001sp}, which studied corrections to the narrow width approximation.}. In this approach, there is no $WWb$ final state, thus the LO contribution to top quark backgrounds comes from the $WWbb$ state (and the $Wt$-$t\bar{t}$ interference is a leading order effect). This contains two gauge-invariant subsets of diagrams containing intermediate top quarks in the narrow-width approximation: (i) singly-resonant diagrams containing one intermediate top quark, such as that shown in figure~\ref{treediags}(a); (ii) doubly-resonant diagrams containing two intermediate top quarks, such as that shown in figure~\ref{treediags}(b). The former could na\"{i}vely be interpreted as (LO) $Wt$ production, and the latter constitute top pair production. However, all interference effects are now included, such that the distinction between $Wt$ and $t\bar{t}$ production is not considered.\\

The resulting calculation for the $WWbb$ final state can be interfaced to a parton shower without worrying about double counting issues, due either to $b$ parton densities (since these are no longer present) or matrix element matching. Regarding the latter, there is no double counting between the shower and the matrix element, because in the four flavor scheme there is no lower order tree level matrix element that, when showered, leads to a $WWbb$ final state (this is not true in the five flavor scheme, in which $WWb$ can shower to give $WWbb$). There are also no further matching issues, due to the lack of a collinear singularity associated with the two final state $b$ quarks. This would be the case even if the $b$ quarks were treated as massless, as in the relevant Feynman diagrams there is never a final state $b$ quark pair resulting from a gluon splitting. The required tree-level matrix elements can be calculated (including full spin correlations in the decay of the top and $W$ bosons) using MadGraph~\cite{Maltoni:2002qb,Alwall:2007st}. We then interface these with HERWIG~\cite{Corcella:2002jc} i.e. the same parton shower that has been used in the MC@NLO results.\\

Having constructed a calculation in which $Wt$ and top pair production are both present inclusive of all necessary interference effects, we now investigate the properties of this description, including its potential accuracy. Our strategy is as follows. We first generate pseudo-data for top production with $t\bar{t}$-like signal cuts, obtained using MC@NLO by combining event samples from $t\bar{t}$ and $Wt$ production. Next, we compare the $WWbb$ description to this, and evaluate the $K$-factor which is necessary to normalize the results of this approach to the pseudo-data. Then we consider $Wt$-like cuts, and see how the $K$-factor needed to normalize the final state analysis to the MC@NLO data compares with the result using $t\bar{t}$ signal cuts. If it is the same, one may argue that it makes sense to model the combination of $Wt$ and $t\bar{t}$ production using a tree level approach normalized to data. If, however, the $K$-factor is not the same for $Wt$-like cuts (or at least similar), this is an argument in favor of separating out $Wt$ and $t\bar{t}$ production as separate production processes in their own right, each with a separate $K$-factor.\\

The above exercise, whilst somewhat academic (since it does not include additional backgrounds due to other single top production modes or non-top related standard model processes) is a useful playground for investigating systematic uncertainty due to interference between $Wt$ and $t\bar{t}$ production. By comparing the results from both calculations, we will be able to discuss and clarify the relative advantages and disadvantages of each approach. In the following section, we discuss the generation of the top pseudo-data.
\subsection{Pseudo-data for top pair production with $t\bar{t}$ selection cuts}
\label{ttbar}
\begin{figure}[h]
\begin{center}
\begin{tabular}{cc}
\scalebox{0.38}{\includegraphics{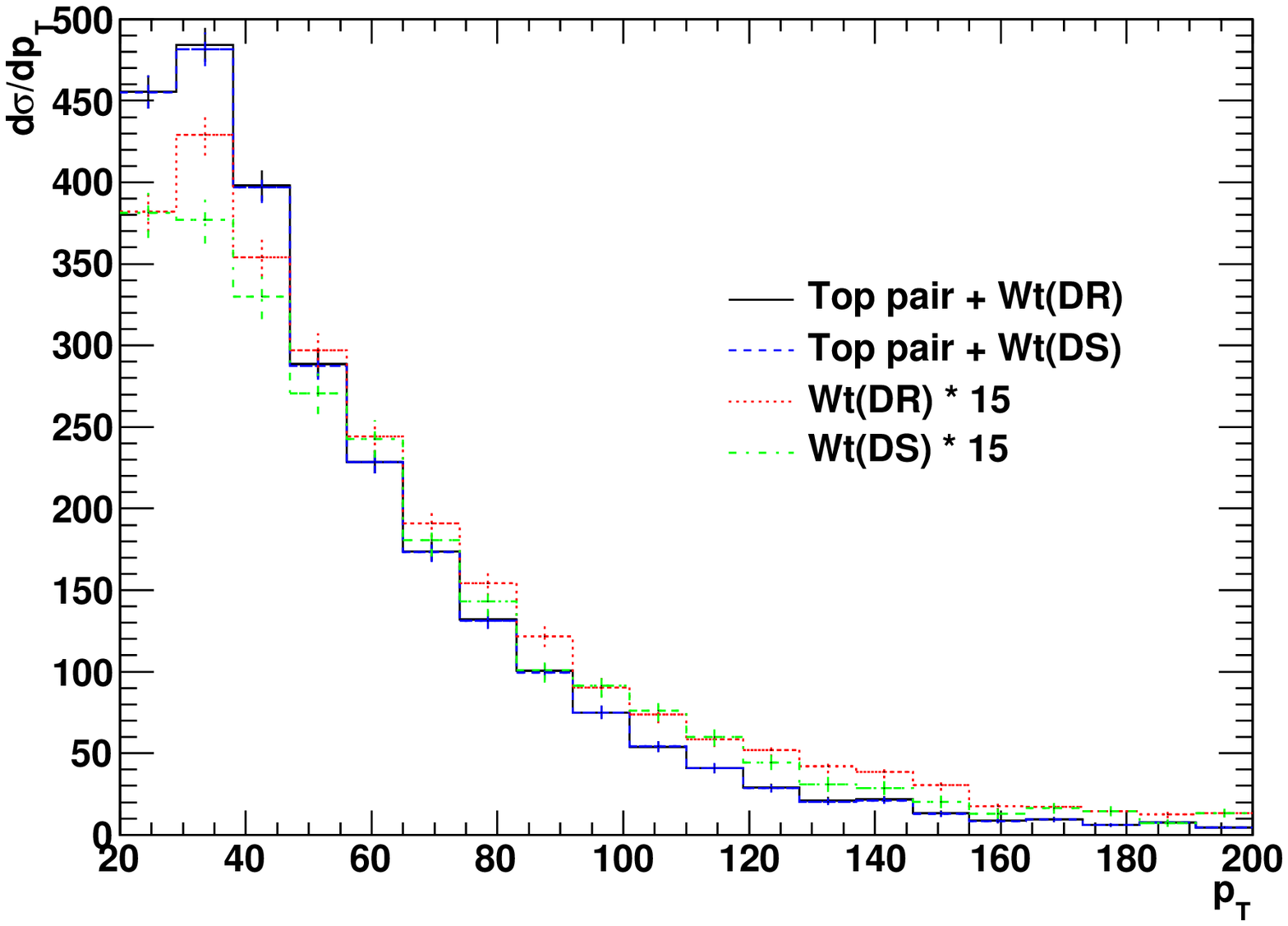}} & \scalebox{0.38}{\includegraphics{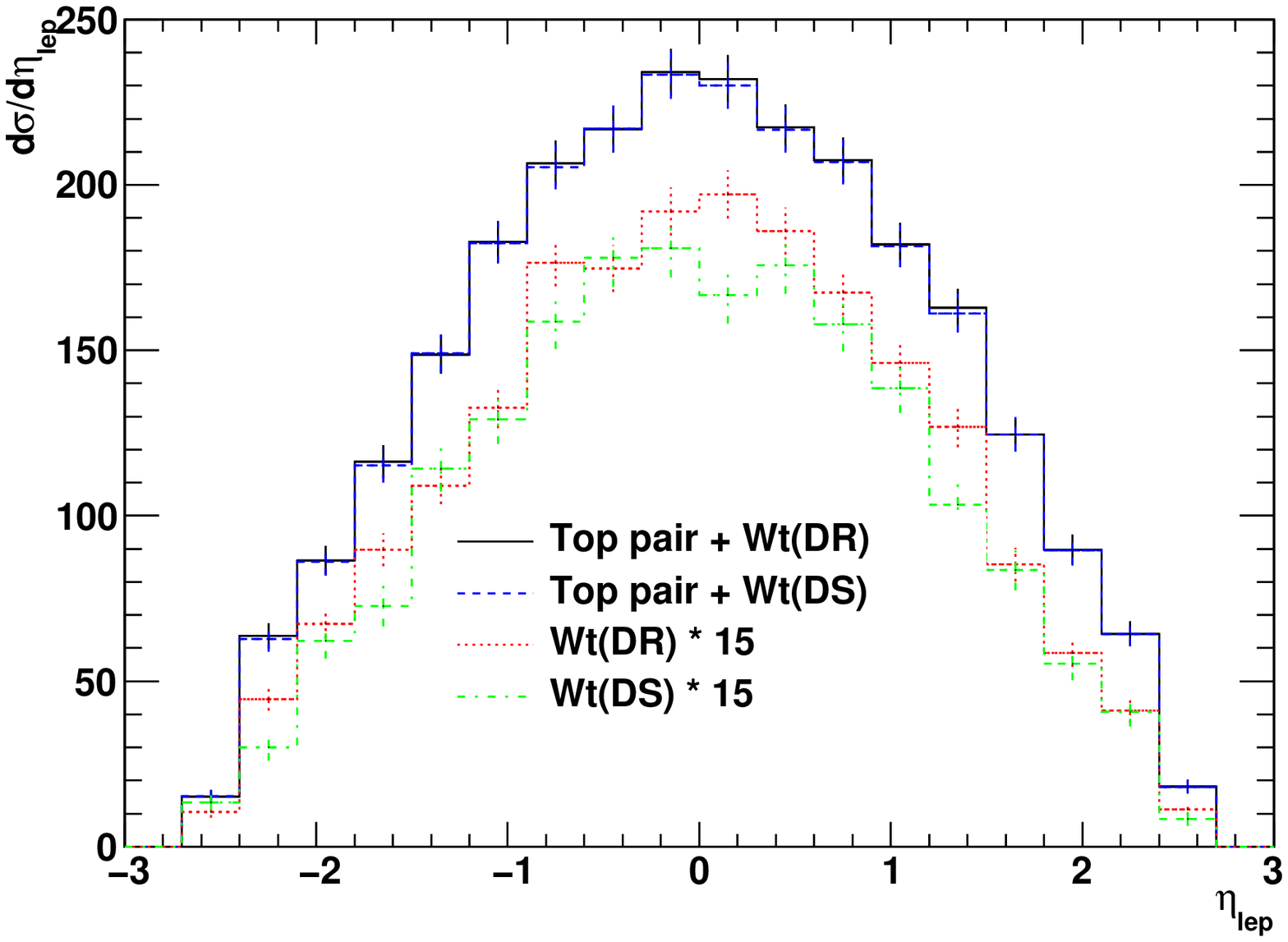}}\\
(a)&(b)
\end{tabular}
\caption{The transverse momentum (a) and pseudo-rapidity (b) distributions of the final state lepton arising from combining MC@NLO event samples for $Wt$ and $t\bar{t}$ production, subject to $t\bar{t}$ signal cuts described in the text. Results are shown for the cases in which the $Wt$ sample is obtained using DR (black), and DS (blue). Uncertainties are statistical, and the vertical axis shows arbitrary units. Also shown are the pure $Wt$ results, multiplied by a constant factor so as to be visible on the same scale.}
\label{ptl_ttbar}
\end{center}
\end{figure}

We form a sample of pseudo-data by running MC@NLO for both the $Wt$ and $t\bar{t}$ production channels, and combining the event samples. We include spin correlations in the decays of the top quarks (and $W$ bosons), and  the $Wt$ results are run using both DR and DS. Parameter choices, parton densities etc. are chosen as in previous sections. Motivated by~\cite{Aad:2009wy}, the following cuts are applied in order to isolate the top pair production cross-section, after requiring semi-leptonic decay of the two $W$ bosons:
\begin{center}
\textbf{$t\bar{t}$ signal cuts}
\end{center}
\begin{enumerate}
\item There must be one lepton (electron or muon) with $p_T>20$ GeV.
\item The missing transverse energy is required to satisfy $E_T^{\text{miss}}>20$ GeV.
\item There must be at least four jets with $p_T>20$ GeV.
\item There must be at least three jets with $p_T>40$ GeV.
\item Leptons and jets must satisfy the pseudo-rapidity cuts $|\eta|<2.5$.
\end{enumerate}
The cross-sections for $Wt$ production and $t\bar{t}$ production are collected in table~\ref{ttbarresults}, together with their total. 
\begin{table}
\begin{center}
\begin{tabular}{c|c}
Process & $\sigma_{\text{NLO}}$/pb\\
\hline
$Wt$ (DR) & $4.27^{+0.3}_{-0.3}$ \\
$Wt$ (DS) & $3.41^{+0.06}_{-0.01}$\\
$t\bar{t}$ & $93.8^{+10}_{-11}$  \\
Total (DR) & $98.1^{+10}_{-11}$\\
Total (DS) & $97.2^{+10}_{-11}$\\
\end{tabular}
\caption{Cross-sections obtained with MC@NLO for $Wt$ and $t\bar{t}$ production, using the top pair production signal cuts of section~\ref{ttbar}. Uncertainties correspond to variation of the common renormalization and factorization scale by a factor of two.}
\label{ttbarresults}
\end{center}
\end{table}
Note that the cuts used to isolate the $t\bar{t}$ signal do not reduce the interference with $Wt$ production, as evidenced by the fact that the DR and DS cross-sections in table~\ref{ttbarresults} differ by around 25\%. However, when combining the event samples, the $t\bar{t}$ component is much larger than the $Wt$ component, so that the systematic uncertainty due to interference between $Wt$ and $t\bar{t}$ has a negligible effect. The two combined cross-sections differ by less than 0.9\%, which is clearly much less than the systematic uncertainty due to scale variation. Furthermore, the total $Wt$ cross-section is less than the scale-variation of the $t\bar{t}$ cross-section. Thus, it is questionable whether $Wt$ production is a significant background at all, let alone whether ambiguities due to interference effects are significant. One may further check that the latter effects are small by comparing kinematic distributions in the two combined event samples. As examples, the transverse momentum and pseudo-rapidity distributions of the final state lepton are shown in figure~\ref{ptl_ttbar}. One sees that indeed the difference between the results for the total of top pair and $Wt$ production is well within statistical uncertainties, although the pure $Wt$ results differ somewhat in shape as well as normalization.\\

In the following subsections, we compare a tree-level (plus parton shower) final state analysis to this pseudo-data. Ideally, one should compare both the MC@NLO and the tree-level approach to real data. Since these are not available, the analysis here allows one, at least to some degree, to compare the relative advantages and disadvantages of each approach. We begin by describing in more detail the tree level calculation.
\subsection{Tree level analysis of final states}
\label{treelevel}
In this section we describe the tree level calculation of the $WWbb$ final state. As explained in the previous section, when considering all diagrams contributing to this final state, one is restricted to a tree-level calculation, as the full NLO amplitudes for production and decay of the relevant top quark intermediate states are not known. Given that the aim of this paper is to address the issue of interference effects in single and double top production, we consider here only those diagrams contributing to the $WWbb$ state that have intermediate top quark resonances (either single or double).\\

Our calculation works as follows. Events are simulated using MadGraph\footnote{Note that MadGraph includes a mass for the $b$ quarks ($m_b=4.7$ GeV) in the hard matrix element, which has not been included in the MC@NLO calculation. We do not expect this to alter our conclusions.} for the process
\begin{equation}
pp\rightarrow W^+W^-b\,\bar{b},
\label{pp}
\end{equation}
where $p$ denotes the proton. As explained in the previous section, initial state $b$ quarks are not present, so as to avoid double counting and matching issues. To be consistent, we use the top quark width as calculated by MadGraph using the masses given above, which is found to be $\Gamma_t=1.407$ GeV. The decay to final state leptons and partons is also present in the MadGraph events, so that spin correlations of decay products are included. In both calculations, the $W$ boson width is set to 2.141GeV, and the branching ratio for semileptonic decays is 24/81. \\

The event output from MadGraph is interfaced to HERWIG, whose parton shower is also used in MC@NLO. The result is then a consistent calculation of the $WWbb$ final state, with both interference and shower effects included. Using default parameters and scales as described previously, the result for the $t\bar{t}$ cross-section is
\begin{equation}
\sigma_{\text{tree}}^{\text{$t\bar{t}$ cuts}}=65.0^{+9.6}_{-11.2}\text{pb},
\label{MGttbar}
\end{equation}
where the superscript $t\bar{t}$ denotes that top pair production signal cuts are applied, rather than that only $t\bar{t}$ intermediate states are considered (which is, of course, not meaningful in this approach). The quoted uncertainty stems from varying the common renormalization and factorization scale by a factor of two, and one sees that this uncertainty is sizeable. From this result and the MC@NLO cross-section given in table~\ref{ttbarresults}, one may define the $K$-factor as the ratio of the central values of the cross-sections\footnote{Note that we use the same parton densities for both the tree level and MC@NLO calculations. This is in contrast to some other definitions of the $K$-factor in which LO and NLO partons are used for LO and NLO calculations respectively. This does not affect our conclusions.}, i.e.
\begin{equation}
K^{\text{$t\bar{t}$ cuts}}_{Wt+t\bar{t}}=\frac{\sigma_{NLO}^{\text{$t\bar{t}$ cuts}}}{\sigma_{\text{tree}}^{\text{$t\bar{t}$ cuts}}}=\left\{\begin{array}{cc}1.508\pm0.012&\text{(DR)}\\1.494\pm0.012&\text{(DS)}\end{array}\right.,
\label{Kfactttbar}
\end{equation}
where the numerator is the MC@NLO combined cross-section for the sum of $Wt$ and $t\bar{t}$ cross-section, obtained using $t\bar{t}$ signal cuts (see section~\ref{ttbar}). Note that the DR and DS results are indistinguishable within statistical uncertainties, as expected from the results of table~\ref{DRvsDS}. \\

\begin{figure}[h]
\begin{center}
\begin{tabular}{cc}
\scalebox{0.38}{\includegraphics{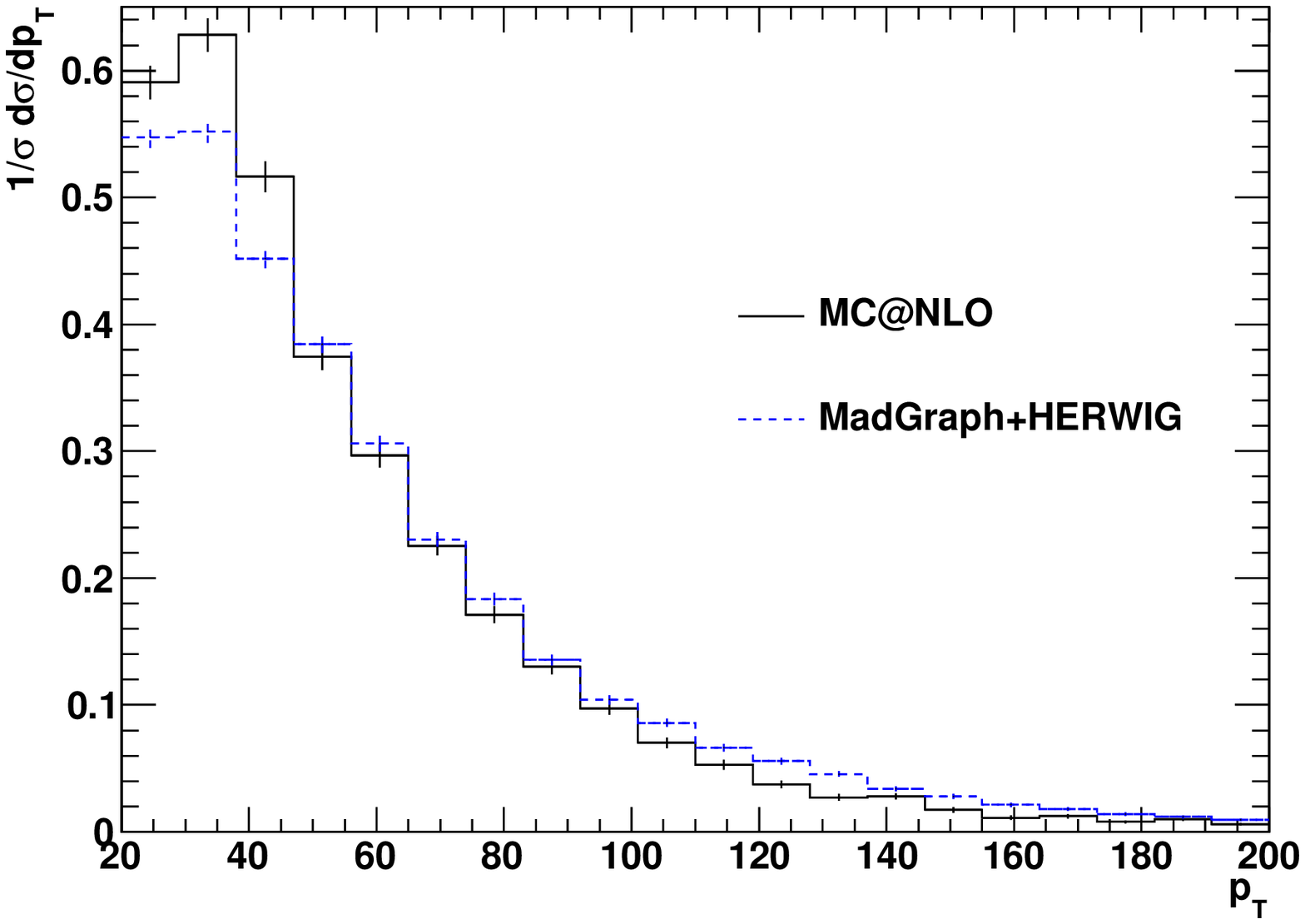}}& \scalebox{0.38}{\includegraphics{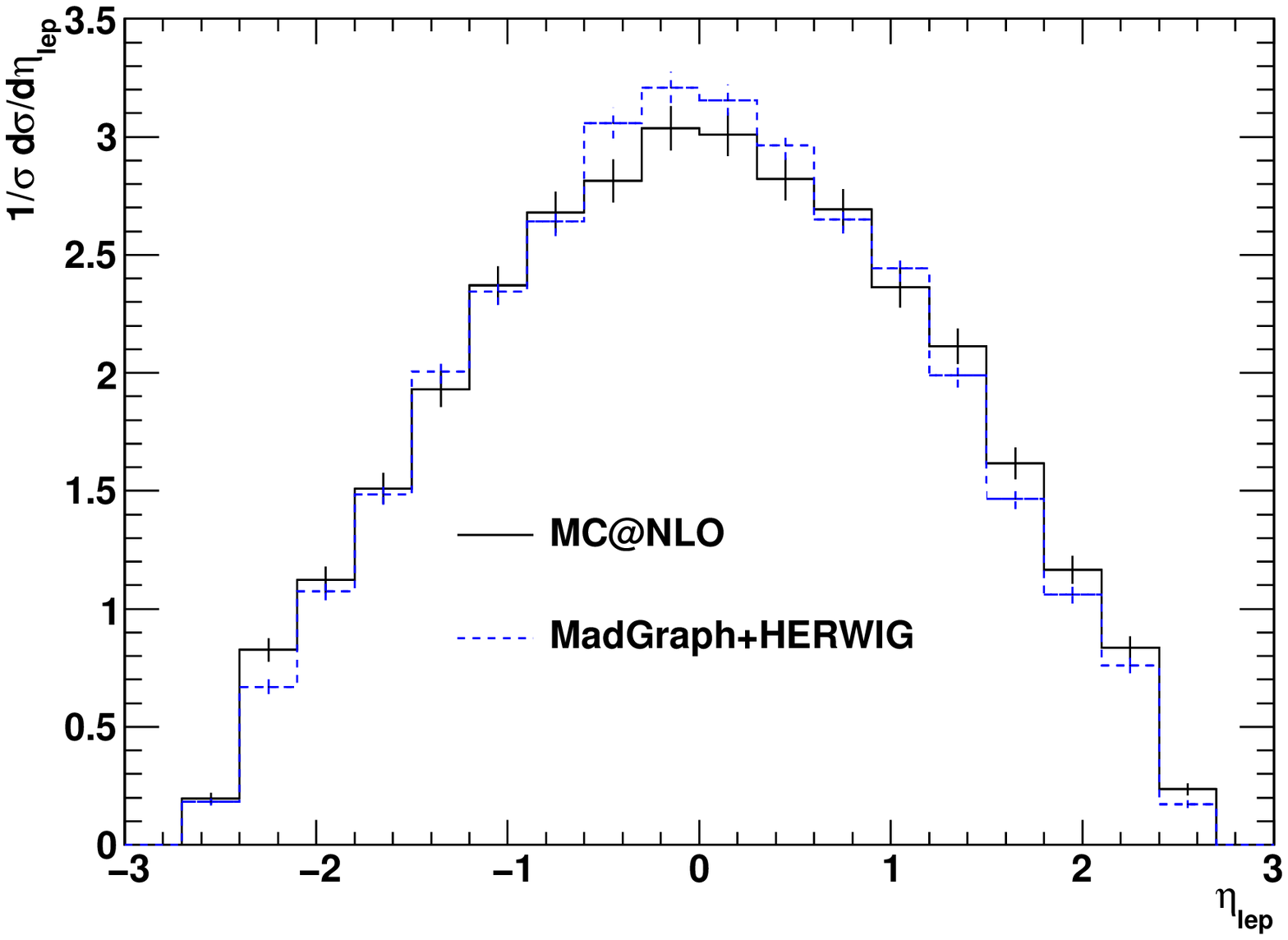}}  \\
(a) & (b)
\end{tabular}
\caption{The transverse momentum (a) and pseudo-rapidity (b) distributions of the final state lepton arising from combining MC@NLO event samples for $Wt$ and $t\bar{t}$ production, subject to $t\bar{t}$ signal cuts described in the text (black). Also shown is the result from the consistent tree level plus parton shower approach discussed in the text. Uncertainties are statistical, and the vertical axis shows arbitrary units.}
\label{ptlLOvsNLO}
\end{center}
\end{figure}

\begin{figure}[h]
\begin{center}
\begin{tabular}{cc}
\scalebox{0.38}{\includegraphics{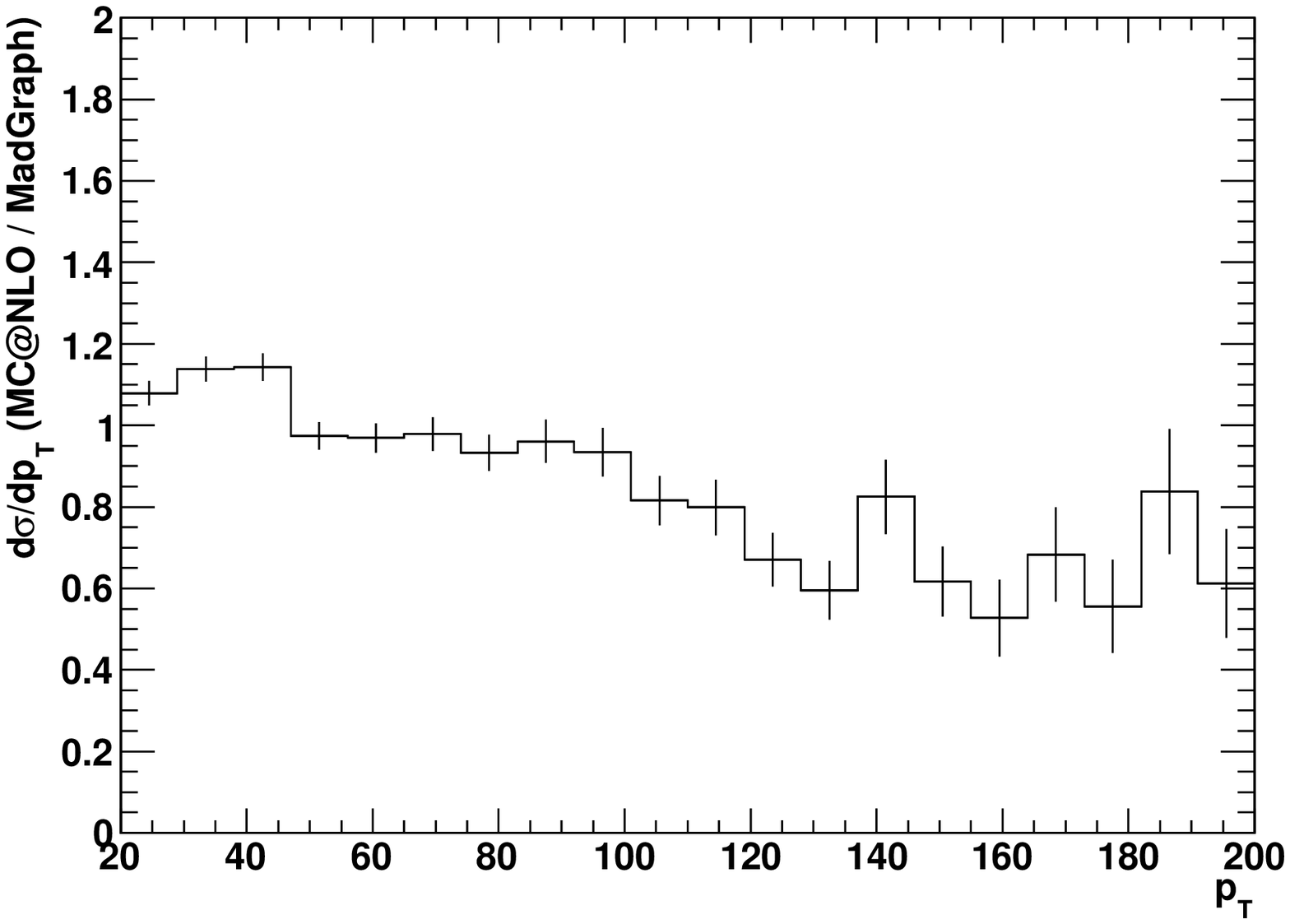}}& \scalebox{0.38}{\includegraphics{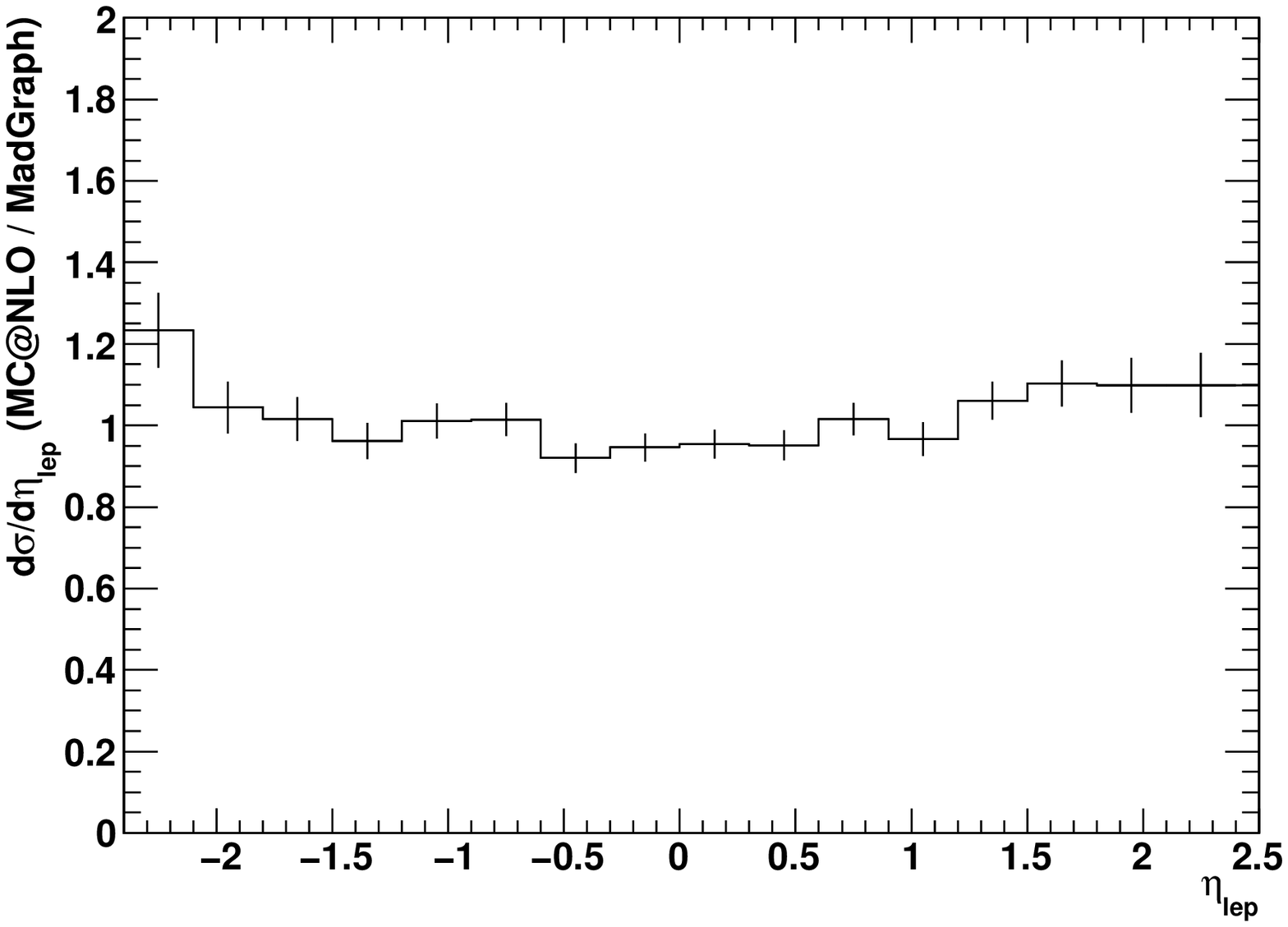}}  \\
(a) & (b)
\end{tabular}
\caption{The ratio of normalized distributions in transverse momentum (a) and pseudo-rapidity (b) of the final state lepton, from the MC@NLO and MadGraph (plus HERWIG) computations, for the top pair production signal cuts discussed in the text. Uncertainties are statistical.}
\label{ptlLOvsNLOrat}
\end{center}
\end{figure}

The lepton transverse momentum and pseudo-rapidity distributions from both the Madgraph and MC@NLO calculations are shown in figure~\ref{ptlLOvsNLO}. One observes some minor difference in shape between the tree level and NLO analyses, which suggests that normalizing the LO results via a $K$-factor is a somewhat limited approximation. The latter can be more clearly seen in figure~\ref{ptlLOvsNLOrat} which shows the ratios, bin by bin, of the leptonic transverse momentum and pseudo-rapidity distributions obtained in both approaches.\\

Having normalized the tree level calculation to the MC@NLO pseudo-data using top pair production signal cuts, one may then investigate what happens for the $Wt$-like signal cuts of section~\ref{signal}. Given that these depend separately on the number of $b$ jets and the number of light jets, the $K$-factor for these cuts (defined analogously to eq.~(\ref{Kfactttbar})) will potentially depend on the $b$-tagging efficiency $e_b$ and the light jet rejection rate $r_{lj}$. Results are shown in table~\ref{WtKfacresults}, where the $K$-factors have been obtained as the ratio of cross-sections from the MC@NLO and MadGraph (plus HERWIG) computations. The former results depend upon whether DR or DS is used for the $Wt$ channel (although we have already seen in section~\ref{signal} that this is a minor effect), thus results are presented for both choices.
\begin{table}
\begin{center}
\begin{tabular}{c|c|c|c}
$e_b$ & $r_{lj}$ & $K^{\text{DR}}$ &$K^{\text{DS}}$\\
\hline
1 & $10^4$ &$1.349\pm 0.024$ & $1.345 \pm 0.028$\\
0.6 & 30 & $1.367 \pm 0.028 $ & $1.362 \pm 0.028 $\\
0.6 & 200 & $1.308 \pm 0.026 $ & $ 1.302\pm 0.026 $\\
0.4 & 300 & $1.357 \pm 0.032$ & $ 1.353\pm 0.032 $\\
0.4 & 2000 &$1.345 \pm 0.032$ & $ 1.342\pm 0.032 $\\
\end{tabular}
\caption{$K$-factors normalizing the tree level $WWbb$ (plus parton shower) calculation to the sum of $Wt$ and $t\bar{t}$ production obtained using MC@NLO, for the $Wt$ signal cuts described in section~\ref{signal}. Results are shown for both DR and DS, and for a range of $b$-tagging efficiencies $e_b$ and light jet rejection rates $r_{lj}$. The quoted uncertainties are statistical.}
\label{WtKfacresults}
\end{center}
\end{table}
From the table, one sees that the $K$-factor does not depend on whether DS or DR is used i.e. the results for each choice of $(e_b,r_{lj})$ are equal within statistical uncertainties. However, the $K$-factor does depend slightly upon the light jet rejection rate $r_{lj}$ and b tagging efficiency $e_b$. \\

\begin{figure}
\begin{center}
\scalebox{0.38}{\includegraphics{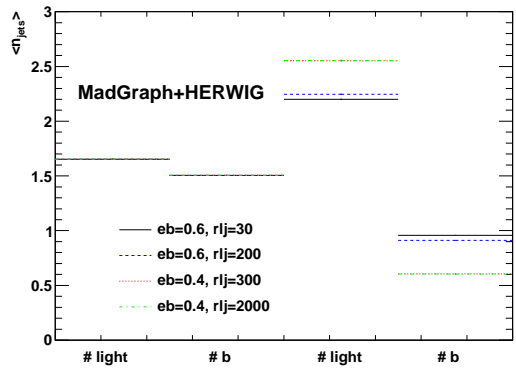}}
\caption{The average number of $b$ and light jets before (two left-most bins) and after (two right-most bins) reshuffling due to $b$-tagging efficiency and light-jet rejection rate. Results are obtained from the MadGraph plus HERWIG calculation, for the $Wt$ signal cuts.}
\label{avgjets_MG}
\end{center}
\end{figure}

One might indeed expect each calculation (i.e. the MC@NLO approach and the tree level plus parton shower analysis) to depend on the $b$-tagging efficiency and/or light jet rejection rate, due to the fact that the cuts involve separate restrictions on the numbers of $b$ and light jets. However, the sensitivity of the $K$-factor to $r_{lj}$ and $e_b$ means that the two calculations are not affected in the same way. This is not surprising, given that the MC@NLO calculation has initial state $b$ quarks whereas the tree level plus parton shower analysis has all $b$ quarks generated from gluon splitting. The hard matrix element in the latter calculation has at least two $b$ quarks in it, whereas the former may have only a single $b$ quark. This, coupled with the requirement of one hard $b$ jet and two light jets in the signal cuts, means that the sensitivity of the two calculations to the light jet rejection rate will be different. That this is not a large effect can be seen by comparing figure~\ref{avgjets_MG} and figure~\ref{avgjets}, which show the average number of $b$ and light jets before and after shuffling in the two approaches. The MadGraph plot of figure~\ref{avgjets_MG} shows that there is not a substantial difference in the number of $b$ or light jets passing the detector cuts between the tree level calculation and the five flavor scheme adopted in MC@NLO. Whether or not one includes initial state $b$ quarks is ultimately a matter of choice, in that both schemes are perturbatively consistent. \\

More significantly, the $K$-factor for the $Wt$ signal cuts is not the same as for the $t\bar{t}$ production cuts but is notably lower (by $\sim15\%$). Note that this difference is significant in the sense that it is larger than the scale variation uncertainty associated with the total $Wt$ plus $t\bar{t}$ cross-section ($\sim 10\%$). That the $K$-factor is lower than that for $t\bar{t}$ signal cuts is not surprising given that previous NLO calculations of the $Wt$ mode~\cite{Campbell:2005bb,Zhu:2002uj} (both of which give some procedure for defining the $Wt$ process) also find that the $K$-factor for pure $Wt$ production is lower than that for $t\bar{t}$. Thus, when signal cuts are used to isolate the $Wt$ signal, one expects that the $K$-factor which normalizes the sum of $Wt$ and $t\bar{t}$ is also reduced.\\

\begin{figure}[h]
\begin{center}
\begin{tabular}{cc}
\scalebox{0.38}{\includegraphics{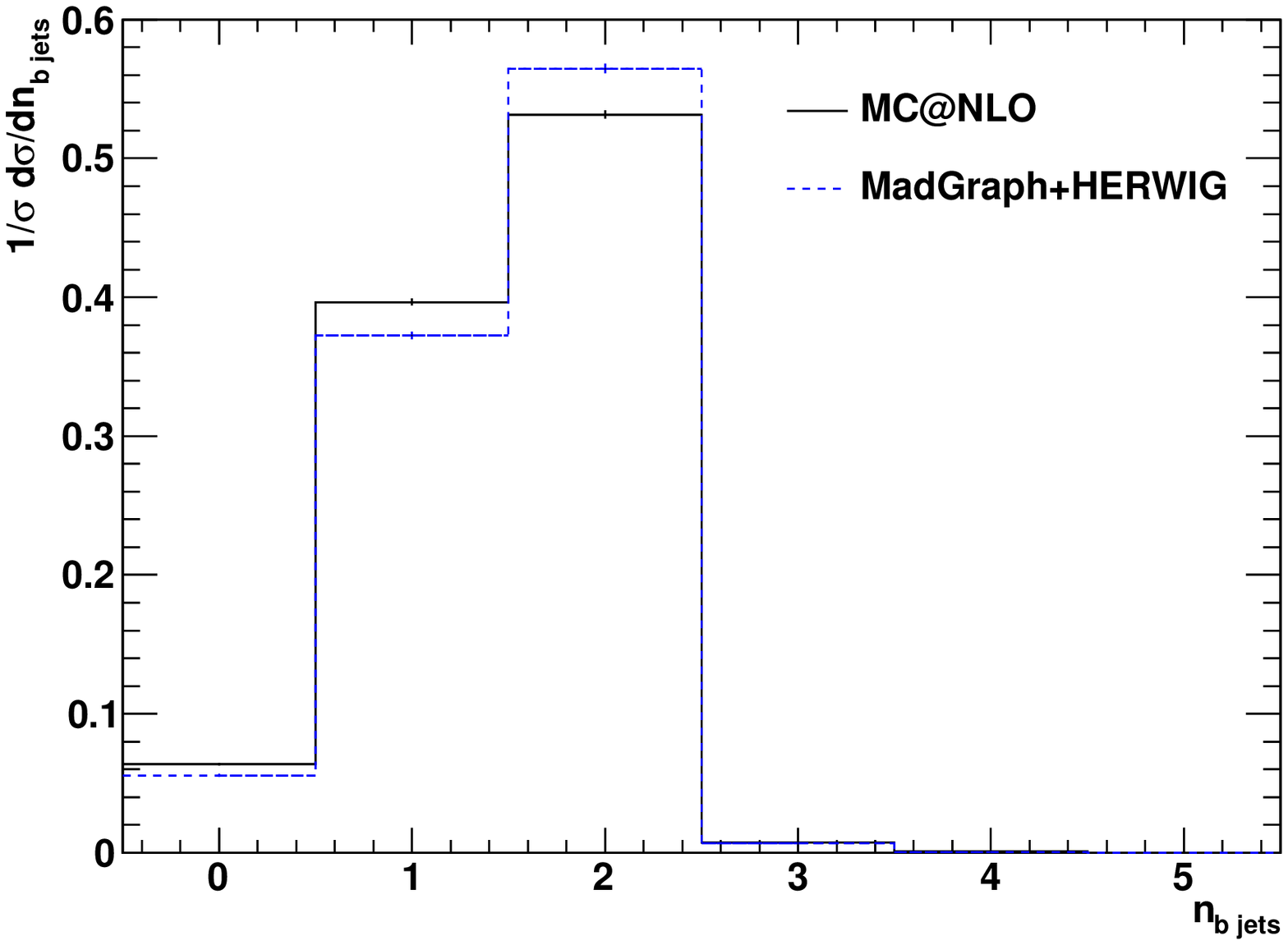}}& \scalebox{0.38}{\includegraphics{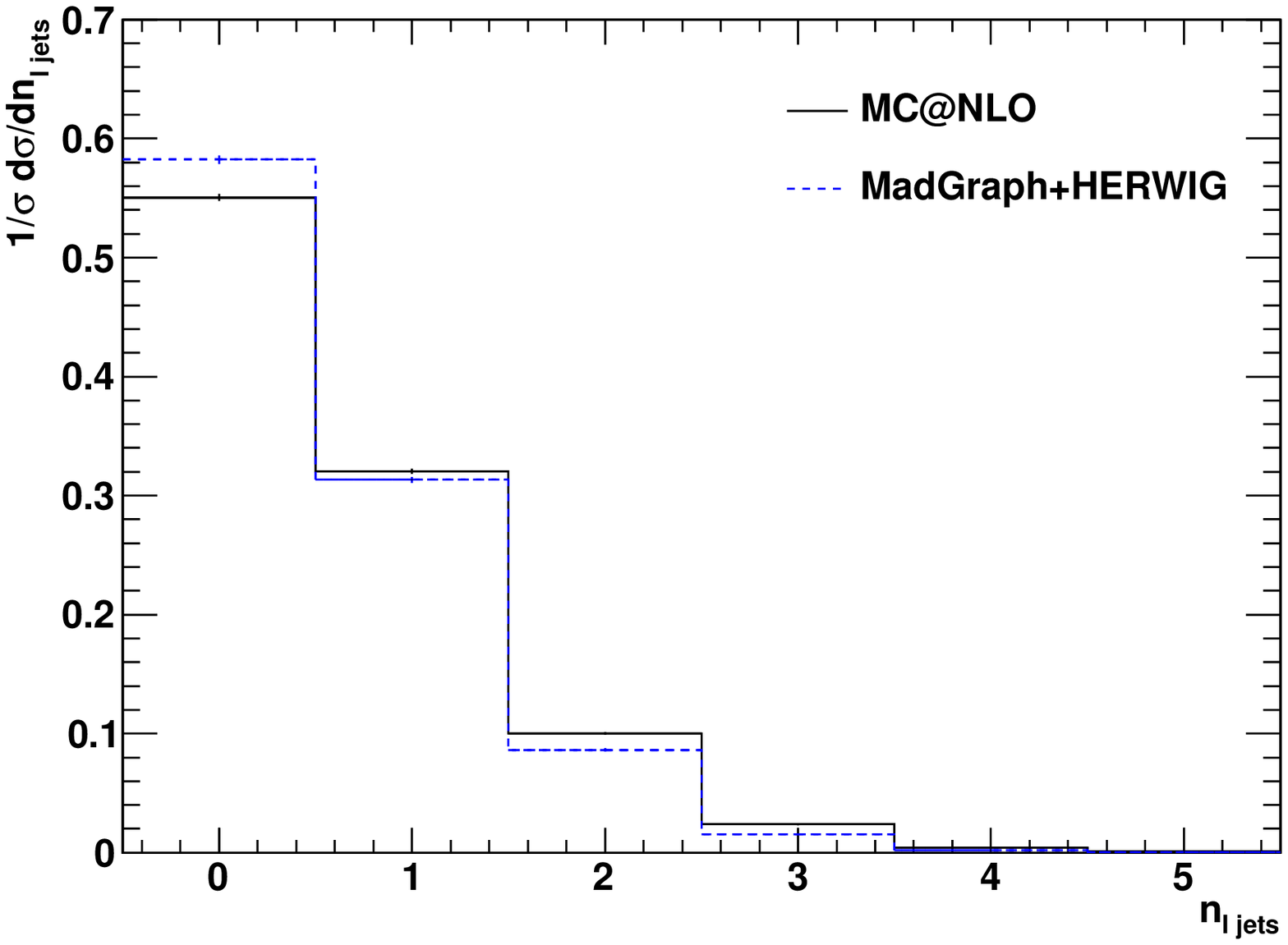}}  \\
(a) & (b)
\end{tabular}
\caption{Distributions of the number of (a) $b$ jets; (b) light jets passing the detector cuts, for the sum of $Wt$ and $t\bar{t}$ production with fully leptonic decays of the $W$ bosons. Results are shown for perfect $b$ tagging efficiency and light jet rejection rate (i.e. $e_b=1$ and $r_{lj}=10^4$).}
\label{higgsjets}
\end{center}
\end{figure}

One may also evaluate a similar $K$-factor for the Higgs signal cuts used in section~\ref{Higgs}. This gives some indication of how well the background to $H\rightarrow WW$ due to top production is estimated, and can be calculated similarly to the result for the $Wt$-like signal cuts. We generate events for the process of eq.~(\ref{pp}), including the leptonic decays of both $W$ bosons, so that spin correlations are included (note that this is particularly important for the Higgs signal cuts, because they include a restriction on the azimuthal angle between the lepton pair). The branching ratio for the leptonic final state is 4/81. Next, the events are interfaced with HERWIG as before, and the $K$-factor is then found to be
\begin{equation}
K^{\text{$H$ cuts}}_{Wt+t\bar{t}}=\frac{\sigma_{NLO}^{\text{$H$ cuts}}}{\sigma_{\text{tree}}^{\text{$H$ cuts}}}=1.98 \pm 0.07,
\label{Kfachiggs}
\end{equation}
where the cross-sections on the right-hand side denote the MC@NLO and Madgraph results for the top production background, and the quoted uncertainty is statistical. The $Wt$ component of the MC@NLO calculation is obtained using diagram removal. Note that the result is higher than the corresponding result for the $t\bar{t}$ cuts, and again is outside the scale variation uncertainty associated with the latter. The former property can be partially explained from the fact that the signal cuts involve a strong veto on any jets passing the detector constraints. Some of the difference in $K$-factor can then be related to the distribution of $b$ and light jets passing the detector cuts in the two calculations (and before additional cuts have been applied). These are shown in figure~\ref{higgsjets}.
The differences between the MadGraph (plus HERWIG) and MC@NLO calculations are as expected. In figure~\ref{higgsjets}(a) one sees that there are less events with no $b$ jets in the MadGraph calculation, presumably due to the fact that a four flavor scheme has been used so that there are always at least two $b$ quarks in the final state. However, there are less events with no light jets in the MC@NLO calculation, due to the fact that the NLO matrix element creates harder light jets on average, which are more likely to pass the detector cuts. These two effects modify the $K$-factor in opposite directions, but the net result is that the MC@NLO calculation has more events with no jets than does the MadGraph calculation -- 3.6\% rather than 3.2\%. \\

To summarize, the above results imply that the MC@NLO description of the sum of the $t\bar{t}$ and $Wt$ cross-sections is not related to the tree level plus parton shower analysis by a simple rescaling. The question then is which is the optimal description, that gives the most accurate comparison to data. The advantage of the tree level analysis is that it consistently combines the $Wt$ and $t\bar{t}$ processes so that any issues regarding the correct inclusion of interference effects are no longer present. However, this would seem to be the only advantage. The MC@NLO approach on the other hand benefits from the usual advantages of combining a NLO matrix element with a parton shower i.e. reduced scale uncertainty, and correct treatment of the first NLO emission. The latter contributes to shape differences in distributions, which have indeed been observed above\footnote{There is also a resummation of logarithms $\sim{\cal O}(\ln(m_t/m_b))$ when a $b$ parton density is used. However, these are not expected to be important, as found in~\cite{Campbell:2009ss}.}. Finally, it is clearly advantageous, given the differences observed above, to have two separate $K$-factors for what are essentially two different processes.
\section{Discussion}
\label{discussion}
In this paper we have addressed the issue of $Wt$ production at the LHC, focussing on whether or not it makes sense to consider this as a production process in its own right. A theoretical ambiguity arises due to interference between $Wt$ and $t\bar{t}$ production, i.e. the same Feynman diagrams contribute to each process. In the five flavor scheme in which a bottom quark parton density is used, this interference occurs at NLO and beyond in $Wt$ (where the relevant diagrams can be interpreted as LO top pair production, with decay of the antitop). Furthermore, in order to test which solutions to this problem are viable in an experimental setting, one must interface the hard matrix element with a parton shower algorithm, necessitating the use of MC@NLO. The problems of implementing $Wt$ production were dealt with in~\cite{Frixione:2008yi}, and the resulting software contains two definitions of the $Wt$ mode such that the difference between them provides a measure of the systematic uncertainty due to interference effects. The aim of this paper has been to extend the results of that paper, by further investigating the circumstances in which such a tool can be used in the context of a realistic analysis. \\

There are two main contexts in which calculation of the $Wt$ mode is necessary. Firstly, there is the isolation of $Wt$ production as a signal, which we considered in section~\ref{signal}. We applied basic cuts designed to isolate this signal, and obtained results using both the DR and DS options in MC@NLO. These were found to give very similar results, agreeing within other systematic uncertainties (e.g. scale variation). Importantly, this agreement persisted in kinematic distributions and for all choices of $b$-tagging efficiency $e_b$. Furthermore, the $Wt$ cross-section was found to be larger than the scale variation associated with the top pair production cross-section (also evaluated using MC@NLO), a feature which is dependent on the choice of signal cuts. Only if the latter property is satisfied is it truly meaningful to address the $Wt$ signal, and that this is indeed the case for fairly primitive cuts is encouraging. This is particularly true given the hope that $Wt$ production can be observed with early LHC data (see e.g.~\cite{Aad:2009wy}), in which case one does not want to have to pay too much of a penalty in the $Wt$ cross-section in order to strengthen the signal to background ratio with respect to top pair production. \\

The second main context in which $Wt$ production occurs is when both this and top pair production are backgrounds to a third process. We considered such a case in section~\ref{Higgs}, where our example signal was Higgs boson production with subsequent decay to a $W$ boson pair. We found that, for the cuts used to isolate this signal, the cross-section for top pair production is comparable with that of $Wt$ production (i.e. within a factor $\simeq 2$). Thus, it is imperative in such a case that $Wt$ production be taken into account. Furthermore, the DR and DS results agreed very well with each other, and certainly well-within scale variation uncertainties. The agreement extended to kinematic distributions, and we showed a couple of examples. The question then remains of whether one has to worry about interference between $Wt$ and $t\bar{t}$ production for other possible signals, and we discussed a number of possibilities. The most general advice that can be given is that if there is any doubt over the validity of separating $Wt$ and $t\bar{t}$ production, a given analysis can be repeated with DR and DS in order to estimate the systematic uncertainty involved. This must then be compared with other uncertainties in order to gauge whether or not the analysis is valid.\\

The possibility remains however of not trying to separate $Wt$ and $t\bar{t}$ production at all, and always attempting to include all Feynman diagrams in a consistent calculation of given final states. We discussed such an approach in section~\ref{finalstates}, in which we interfaced a tree level calculation of the $WWbb$ final state (in which all initial state $b$ quarks were generated via gluon splittings) with the HERWIG parton shower algorithm i.e. the same parton shower that is used in MC@NLO. We normalized this calculation to the MC@NLO results for cuts used to isolate the top pair production signal. We then evaluated corresponding $K$-factors for $Wt$ signal cuts, and found that the factor needed was different to that obtained for the top pair production cuts, indicating that one calculation is not a straightforward rescaling of the other. This was further confirmed by the $K$ factor for the Higgs signal cuts, which was different again, and large ($\simeq 2$). These results are not surprising, given the difference between the two approaches, and raise the question of which is the right approach to adopt. One could claim of course that the MC@NLO calculation, in neglecting interference effects, is flawed. Or, that the estimate of systematic uncertainty provided by the DR and DS codes is not a good estimate, however this can be obtained. We believe that such a viewpoint is unduly pessimistic, for several reasons.\\

Firstly, the fact that the MC@NLO approach neglects interference diagrams (i.e. diagrams with a top pair intermediate state, where the invariant mass of the antitop is far off-shell), whilst an approximation, seems to be a very good approximation throughout much of the phase space. The evidence is presented, through numerous examples of total cross-sections and kinematic distributions, in this paper. Furthermore, {\it any} fixed order calculation is an approximation to the underlying physics, and one must carefully consider of a number of alternatives which gives the best approximation. The tree level approach described above, whilst a consistent combination of Feynman diagrams, suffers from a large scale uncertainty, as is typical of LO calculations. Given also the fact that the $K$-factors for the two sets of cuts also differ outside this uncertainty, it seems natural to concede that MC@NLO provides a better approximation of the underlying physics than the tree level calculation matched to a parton shower.  \\

Such a conclusion is fortunate also for practical and technical reasons. It is clearly better, if $Wt$ and $t\bar{t}$ can be separated, to have the possibility to normalize each separately to data. This allows greater flexibility in estimating the top quark backgrounds to other processes. Furthermore, in searching for single top production it is useful to have a means of efficiently generating events which pass $Wt$-like signal cuts. MC@NLO provides a solution to this problem, in that it cleanly separates $Wt$ and $t\bar{t}$ production as far as running is concerned.\\

To conclude, we have critically examined whether one can separate $Wt$ and $t\bar{t}$ production in a number of contexts. It seems perfectly possible to try to isolate $Wt$ production as a signal at the LHC, and existing LO analyses can be profitably generalized to NLO using MC@NLO. 
\section{Acknowledgements}
\label{acks}
CDW is supported by a Marie Curie Intra-European Fellowship, entitled ``Top Physics at the LHC'', and is grateful to Rikkert Frederix for many helpful discussions regarding MadGraph. EL is supported by the Netherlands Foundation for Fundamental Research of Matter (FOM) and the National Organization for Scientific Research (NWO). FM is partially funded by Technical and Cultural Affairs through the
Interuniversity Attraction Pole P6/11. CDW, EL and FM are grateful to the CERN theory group for hospitality.

\bibliographystyle{JHEP}
\bibliography{refs}
\end{document}